\PassOptionsToPackage{numbers,sort&compress}{natbib}  
\documentclass[preprint,12pt]{elsarticle}
\usepackage[margin=2.5cm]{geometry}
\usepackage{enumitem}
\newlist{inlinelist}{enumerate*}{1}
\setlist*[inlinelist,1]{%
  label=(\roman*),
}
\usepackage{soul}
\usepackage{hhline}
\usepackage{subfig}
\usepackage{multirow}
\usepackage{xcolor}
\usepackage{graphicx}
\usepackage{upgreek}
\usepackage{amssymb}
\usepackage{amsfonts,amsthm,bm,amsmath} 
\usepackage{appendix}
\usepackage{times}
\usepackage{caption}
\usepackage{subfig}
\newcommand{\psubref}[1]{\protect\subref{#1}}
\usepackage{multirow}
\usepackage{xcolor}
\usepackage[linesnumbered,ruled,vlined]{algorithm2e}
\usepackage{tablefootnote}

\SetCommentSty{mycommfont}

\SetKwInput{KwInput}{Input}                
\SetKwInput{KwOutput}{Output}              

\usepackage[colorlinks]{hyperref} 
\hypersetup{ 
    colorlinks=true,       
    linkcolor=red,          
    citecolor=blue,        
    filecolor=magenta,      
    urlcolor=cyan           
}

\newcommand{\fref}[1]{Fig.~\ref{#1}}

\newcommand{\sref}[1]{Section~\ref{#1}}
\newcommand{\tref}[1]{Table~\ref{#1}}
\setcitestyle{square}

\journal{Acta Mechanica}
\begin{document}

\begin{frontmatter}

\title{Predictions of Transient Vector Solution Fields with Sequential Deep Operator Network}
\author[]{Junyan He$^{1}$}
\author[]{Shashank Kushwaha$^{1}$}
\author[]{Jaewan Park$^{1}$}
\author[]{Seid Koric$^{1,2}$\corref{mycorrespondingauthor}}
\cortext[mycorrespondingauthor]{Corresponding author}
\ead{koric@illinois.edu}
\author[]{Diab Abueidda$^{2}$\corref{mycorrespondingauthor2}}
\cortext[mycorrespondingauthor2]{Corresponding author}
\ead{abueidd2@illinois.edu}
\author[]{Iwona Jasiuk$^1$}

\address{$^1$ Department of Mechanical Science and Engineering, University of Illinois at Urbana-Champaign, Urbana, IL, USA \\
$^2$ National Center for Supercomputing Applications, University of Illinois at Urbana-Champaign, Urbana, IL, USA \\
}

\begin{abstract}
The Deep Operator Network (DeepONet) structure has shown great potential in approximating complex solution operators with low generalization errors. Recently, a sequential DeepONet (S-DeepONet) was proposed to use sequential learning models in the branch of DeepONet to predict final solutions given time-dependent inputs. In the current work, the S-DeepONet architecture is extended by modifying the information combination mechanism between the branch and trunk networks to simultaneously predict vector solutions with multiple components at multiple time steps of the evolution history, which is the first in the literature using DeepONets. Two example problems, one on transient fluid flow and the other on path-dependent plastic loading, were shown to demonstrate the capabilities of the model to handle different physics problems. The use of a trained S-DeepONet model in inverse parameter identification via the genetic algorithm is shown to demonstrate the application of the model. In almost all cases, the trained model achieved an $R^2$ value of above 0.99 and a relative $L_2$ error of less than 10\% with only 3200 training data points, indicating superior accuracy. The vector S-DeepONet model, having only 0.4\% more parameters than a scalar model, can predict two output components simultaneously at an accuracy similar to the two independently trained scalar models with a 20.8\% faster training time. The S-DeepONet inference is at least three orders of magnitude faster than direct numerical simulations, and inverse parameter identifications using the trained model is highly efficient and accurate. 
\end{abstract}


\end{frontmatter}

\section{Introduction}

\label{sec:intro}
The recent advancements in high-performance computing and machine learning (ML) techniques has led to significant strides in the neural network (NN) applications in various fields, including structural optimization \citep{bastek2023inverse,sosnovik2019neural}, flow prediction \citep{lira2022computational,belbute2020combining,ye2023locality}, additive manufacturing \citep{valizadeh2022convolutional,kwon2020deep}, and exploring structure-property relations \citep{kushwaha2023designing, he2023exploring}. Besides approximating the underlying physics via simulation data (data-driven NNs), various physics-informed NNs have been proposed to embed the physics principles directly into the loss function, thereby aleviating the need for simulation data \citep{fuhg2022mixed,he2023deep,nguyen2020deep,nguyen2021parametric}. Trained NNs have been successfully used as efficient surrogate models in inverse design \citep{bastek2023inverse,liu2018training}, design optimization \citep{cook2000combining,wang2005hybrid}. For a comprehensive overview of the application of NNs in computational mechanics, see \citep{Herrmann2024}.

NNs can be trained to approximate a particular solution, or it can be trained to approximate the underlying physics/mathematical operator for a class of problems in what is known as operator learning \citep{kovachki2023neural}. Previous researchers have proposed different architectures for operator learning, with two notable ones being the Fourier neural operator (FNO) and the Deep Operator Network (DeepONet). FNO was first proposed by Li et al. \citep{li2020fourier} to solve partial differential equations with parametric inputs. Inspired by the Fourier transform used in solving differential equations, the input function goes through multiple Fourier layers. The encoded information is then mapped onto the output function space. Each Fourier layer takes the Fast Fourier Transform (FFT) of its input and filters out high frequency modes. FNO and its improved versions have been successfully applied to Burger's equation, Darcy flow, Navier-Stokes equation \citep{li2020fourier}, as well as in elasticity and plasticity problems in the solid mechanics field \citep{li2022fourier}. Modified versions of FNOs were widely used in various areas such as material modeling \citep{you2022learning}, seismic wave equations \citep{li2023solving}, and 3D turbulence for large eddy simulation \citep{li2022eddy}. However, since FNO uses FFT, it suffers difficulties when handling complex geometries or intricate non-periodic boundary conditions. Also, Fourier-based operations are computationally expensive when dealing with high-dimensional problems. Lately, the DeepONet proposed Lu et al. \citep{lu2021learning} emerged as another capable architecture for operator learning. The neural network architecture learns an operator from the governing differential equation of the system, which enables retraining not necessary despite the input to the system, such as parametric functions, boundary conditions, or even varying geometry. DeepONet consists of two separate neural networks, named branch and trunk. The branch network takes the parametric function as an input, while the trunk network takes domain geometry. Both the branch and trunk were fully connected neural networks (FNNs) in the very first form suggested by Lu et al. \citep{lu2021learning}, encoding their input information each, respectively. The output of the branch and trunk were then underwent a dot product to map the parametric functions to the target, the solution to the governing differential equation. Koric et al. \citep{koric2023deep} successfully predicted stress field on a 2D domain with a plastic deformation. Moreover, DeepONets lately have been frequently applied to solve science and engineering problems such as inverse designing nanoscale heat transport system \citep{lu2022multifidelity}, predicting elastic-plastic stress on convoluted topology-optimized domain \citep{HE2023116277}, earthquake localization by Haghighat et el. \citep{haghighat2023novel}, heat conduction problem by Koric and Abueidda \citep{koric2023data}, and nonequilibrium thermodynamics of chemical mixtures combined with physics-informed neural networks by Li et al. \citep{li2023phase}. A comprehensive comparison between FNO and DeepONet was performed by Lu et al. \citep{lu2022comprehensive}

Many real-world engineering problems involve dynamic, time-dependent loading conditions caused by impact, vibrations or cyclic loading, resulting in a time-dependent response in the system. Often times the full flow field (for CFD) or the stress field (for solid mechanics) is needed to gain critical insights in different local regions. Using classical techniques, including finite element analysis (FEA), topology optimization (TO), and sensitivity analysis to obtain a time-dependent response is computationally expensive and time-consuming. Hence, a scientific need exists for more robust data-driven surrogate models for time-dependent problems capable of predicting full-field contours for multiple time steps and vector components. Recurrent neural network (RNN) models like long short-term memory (LSTM) \citep{schmidhuber1997long} and gated recurrent unit (GRU) \citep{cho2014learning} are typically used to capture the time-dependent and causal relationship in the inputs and outputs. However, RNNs are typically used to predict a 1D time sequence instead of a full-field contour. Realizing the effectiveness of the DeepONet in predicting full-field solution contour with parametric inputs, He et al. \citep{HE2024107258} proposed a modification of the DeepONet model (termed S-DeepONet) to combine the temporal encoding capability of RNNs with the spatial encoding capabilities of DeepONet to predict outcomes for the last time step, particularly for scalar output. In this work, we proposed an improvement over the original S-DeepONet architecture to simultaneously predict the full-field solution at different time steps as well as for different vector components, addressing the need to predict time-dependent vector fields in a single model. In this work, we tested our approach with two example problems: (1) a lid-driven cavity flow and (2) a path-dependent plasticity problem, both involving time dependence and representing real-world engineering use cases. We used the proposed innovative S-DeepONet formulation to predict full-field solutions and compared performance with the classical DeepONet formulation. Furthermore, we demonstrate the use of the trained NN in an inverse parameter identification via genetic algorithm.

This paper is organized as follows: \sref{sec:methods} introduces neural network architectures and provides details on the data generation method. \sref{sec:results} presents and discusses the performance of the NN model. \sref{sec:conc} summarizes the outcomes and limitations and highlights future works.

\section{Methods}
\label{sec:methods}
\subsection{S-DeepONet with multiple output dimensions}
\label{sec:NN}
The key idea in the original S-DeepONet architecture as proposed by the authors \cite{HE2024107258} is the separation of the temporal component (handled by a GRU branch network) and the spatial component (handled by a FNN trunk network) of the solution operator, and combining the components via a matrix-vector product and a bias:
\begin{equation}
    \bm{G}_{n} = \sum_{h=1}^{HD} \bm{B}_{h} \bm{T}_{nh} + \beta,
\end{equation}
where $\bm{G}$, $\bm{B}$ and $\bm{T}$ denote the outputs of the S-DeepONet, branch network, and trunk network, respectively. Dimension index $h$ represents the hidden dimension (HD) of branch and trunk networks, and index $n$ represents the flattened spatial dimension, which contains $N$ nodes in the simulation domain. Finally, $\beta$ is a bias added to the product. This structure allows the prediction of a full-field solution discretized by $N$ nodes, where the time-dependent input load information is encoded in the branch network and the spatial geometry information is encoded in the trunk network.

\begin{figure}[h!] 
    \centering
         \includegraphics[width=0.95\textwidth]{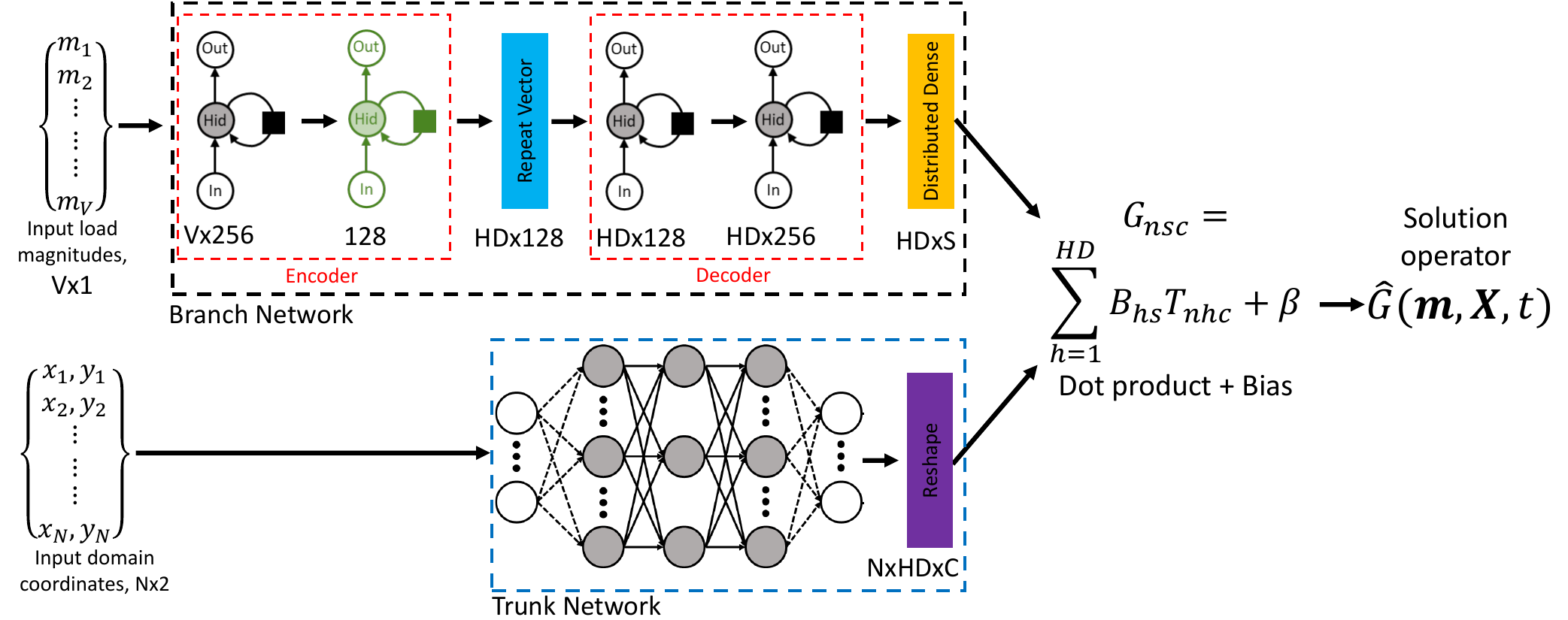}
    \caption{Schematic of the improved S-DeepONet used in this work. $m_V$, $x$, $y$ and $\Hat{G}$ denote the load magnitude, X coordinates, Y coordinates, and the approximated solution operator. Dimensions $HD$, $S$, $N$ and $C$ are the number of hidden dimensions, time steps, nodes in the domain and output vector components, respectively. The black GRU blocks return a sequence (2D outputs), while the green GRU block compresses the output into 1D.}
    \label{architecture}
\end{figure}

\begin{figure}[h!] 
    \centering
         \includegraphics[trim={0cm 2.6cm 0cm 0.cm},clip,width=0.9\textwidth]{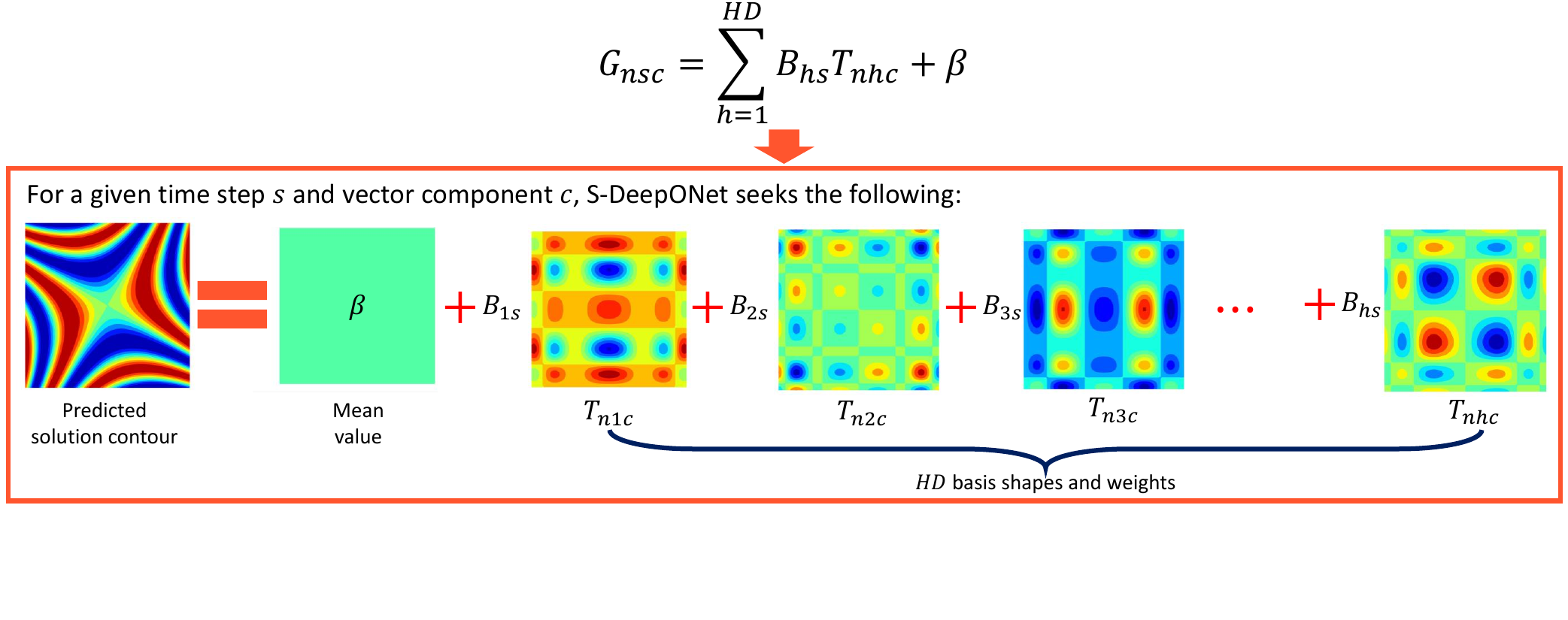}
    \caption{Basis and weight interpretation of the S-DeepONet combination operator. For each vector component $c$, S-DeepONet identifies $HD$ basis shapes from the trunk network. Simultaneously, for each time step $s$, $HD$ weights from the branch network. The solution contour is expressed as the linear combination of all basis shapes including the data mean.}
    \label{basis}
\end{figure}

Although including time-dependent information in the input load, the original S-DeepONet only predicts a scalar solution field at the end of the load. To extend the S-DeepONet architecture to predict vector solution fields at different time steps, we further exploit the idea of separation of spatial and temporal components into the trunk and branch networks. To this end, a novel S-DeepONet structure is proposed, and its schematic is shown in \fref{architecture}.
In the proposed architecture, we leverage the GRU branch network to produce encoded hidden outputs for all the output time steps as a tensor $\bm{B}$ of shape $[HD \times S]$, where $S$ is the number of time steps. Similarly, when provided an $[N \times 2]$ input vector containing 2D coordinates of all nodes in the domain, the trunk network in the proposed S-DeepONet produces an encoded output tensor $\bm{T}$ of shape $[N \times HD \times C]$, where $C$ is the number of output vector components. $\bm{T}$ contains the encoded hidden outputs for all nodes and all vector components. To account for the new output dimensions, we combine the information from the branch and trunk via the following product:
\begin{equation}
    \bm{G}_{nsc} = \sum_{h=1}^{HD} \bm{B}_{hs} \bm{T}_{nhc} + \beta,
\end{equation}
where the lower-case indices $s$ and $c$ correspond to the time step and output component dimensions. The tensor product nature of this combination allows for efficient simultaneous generation of full-field vector outputs at multiple output time steps. This combination can be understood as a simultaneous identification of a set of "basis" shapes (from trunk network) and corresponding weights (from branch network), where the final solution contours are expressed as the weighted linear combination of those basis contours. This idea of basis and weight identification of the S-DeepONet architecture is illustrated further in \fref{basis}.
By predicting $C$ output vector components together, we are eliminating the need to train $C$ different single-component S-DeepONets, each specialized in one output component, thus leading to tremendous computational time savings.

The branch network of the developed S-DeepONet consists of four GRU layers. The encoding-decoding structure of the original S-DeepONet is adopted, with the first two being the encoder and the last two being the decoder. All GRU layers use a tanh activation function. Finally, a time-distributed dense layer with linear activation is used to output the branch results with a hidden dimension ($HD$) of 32 for all $S$ time steps. The trunk network is a FNN with the following seven layers of neurons: $[2, 101, 101, 101, 101, 101, HD \times C]$, where $C$ takes different values in the numerical examples presented in this work. All NNs were implemented in the DeepXDE framework \cite{lu2021deepxde} with a TensorFlow backend \cite{tensorflow2015-whitepaper}. All models were trained for 300000 epochs with a batch size of 64. The Adam optimizer \cite{kingma2014adam} was used and the scaled mean squared error (MSE) was used as the loss function, which is defined as:
\begin{equation}
    {\rm{MSE}} = \frac{ 1 }{ N } \sum^N_{i=1} (f_{FE} - f_{Pred})^2,
\end{equation}
where $N$, $f_{FE}$, and $f_{Pred}$ denote the number of data points, the FE-simulated field value, and the NN-predicted field value, respectively.

\subsection{Data generation}
\label{sec:data_gen}
In this work, we demonstrate the application of the proposed improved S-DeepONet in time-dependent computational fluid dynamics (CFD) and history-dependent plastic deformation problems. Using the trained S-DeepONet, inverse parameter identification is performed with genetic algorithm.

\subsubsection{Lid-driven cavity flow}
\label{sec:cfd}
In the first example, we consider a classical CFD benchmark problem: the lid-driven cavity flow. Consider a rectangular cavity of length 3 and height 1, which is discretized into a $121 \times 41$ uniform grid with 4961 nodes. A lid is located at $y=1$ and moves at a time-dependent velocity profile $\Bar{u}(t)$ in the $X$ direction, driving the fluid motion in the cavity. The radial basis interpolation (RBI) was used to generate the smooth lid velocity profile, which are defined by six uniformly spaced control points. The starting velocity (i.e., at $t=0$) is 0, and the velocity in all other control points were sampled from the range $[-2,2]$. The fluid is assumed to be incompressible, and the system is governed by the Navier-Stokes equation:
\begin{equation}
    \frac{ \partial \bm{u} }{ \partial t } = \mu \nabla^2 \bm{u} - ( \bm{u} \cdot \nabla ) \bm{u} - \frac{1}{\rho} \nabla P,
\end{equation}
where $\bm{u}$, $\mu$, $\rho$, and $P$ denote the velocity vector, viscosity, mass density, and pressure, respectively. In this example, a hypothetical fluid with $\rho=1$ and $\mu=0.1$ was used. In 2D, let the $X$ and $Y$ components of the velocity vector be denoted as $u$ and $v$. No-slip boundary condition was used for all four boundaries:
\begin{equation}
\begin{aligned}
    u = \Bar{u}(t), \, v = 0 \;\; \rm{if} \; y = 1,\\
    \bm{ u } = \bm{0}, \;\; \rm{Otherwise}.\\
\end{aligned}
\end{equation}
For the pressure degree of freedom, the following boundary conditions were used:
\begin{equation}
\begin{aligned}
    P = 0 \;\; \rm{if} \; y = 1,\\
    \frac{\partial P}{\partial x} = 0 \;\; \rm{if} \; x = 0,3,\\
    \frac{\partial P}{\partial y} = 0 \;\; \rm{if} \; y = 0.\\
\end{aligned}
\end{equation}
The trivial initial conditions $\bm{u}=\bm{0}$ and $P=0$ were used. For this simple geometry, the second-order central finite difference (FD) method was used to discretize the governing equation, and the pressure-projection method with explicit time integration was used to evolve the system for 10000 time steps with a time step size of $2 \times 10^{-4}$. An Python code was used to solve this problem, which was adopted from the work of Barba et al. \cite{barba2018cfd}. A total of 4000 simulations were conducted with distinct lid velocity profiles. The primary variables $P$, $u$, and $v$ were stored at 25 uniformly spaced output time steps over the simulation period. For better NN training and accuracy, it is best to scale the input and output data to suitable ranges. To account for the change in solution scale as the fluid flow develops, a time-dependent scale factor for each output variable was first calculated. For a time step $i$, this scale is defined as the maximum absolute value of the field values (over all cases and all points) at this time step. The data at time step $i$ was then divided by this scale, leading to a new data scale of $[-1,1]$.

\subsubsection{Dog-bone axial loading}
\label{sec:plastic_def}
\begin{figure}[h!] 
    \centering
     \subfloat[]{
         \includegraphics[trim={0cm 0cm 0cm 0.cm},clip,width=0.6\textwidth]{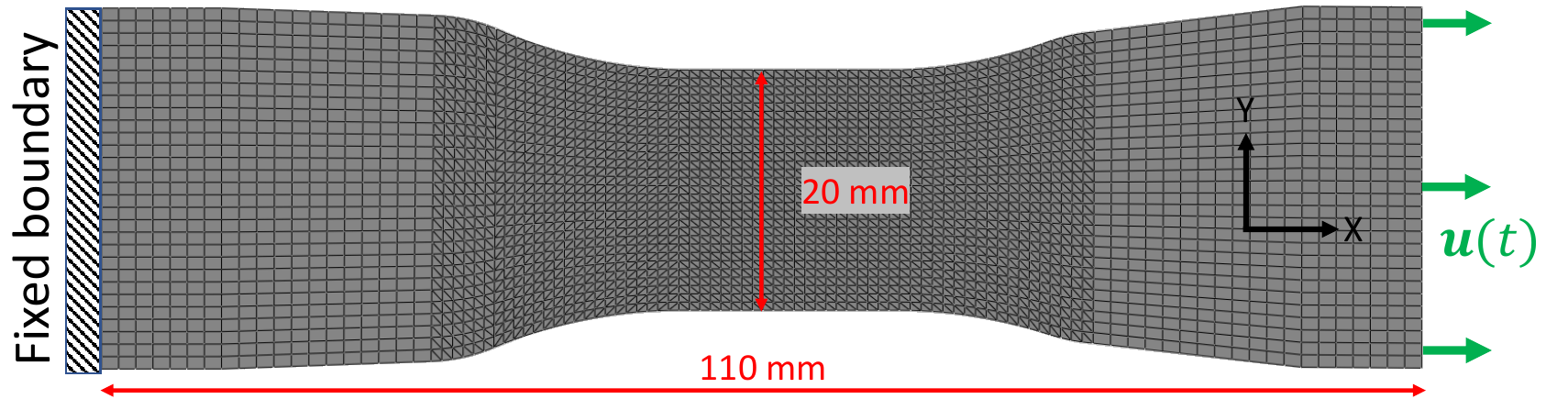}
         \label{geom}
     }
     \subfloat[]{
         \includegraphics[trim={0cm 0cm 0cm 0.cm},clip,width=0.35\textwidth]{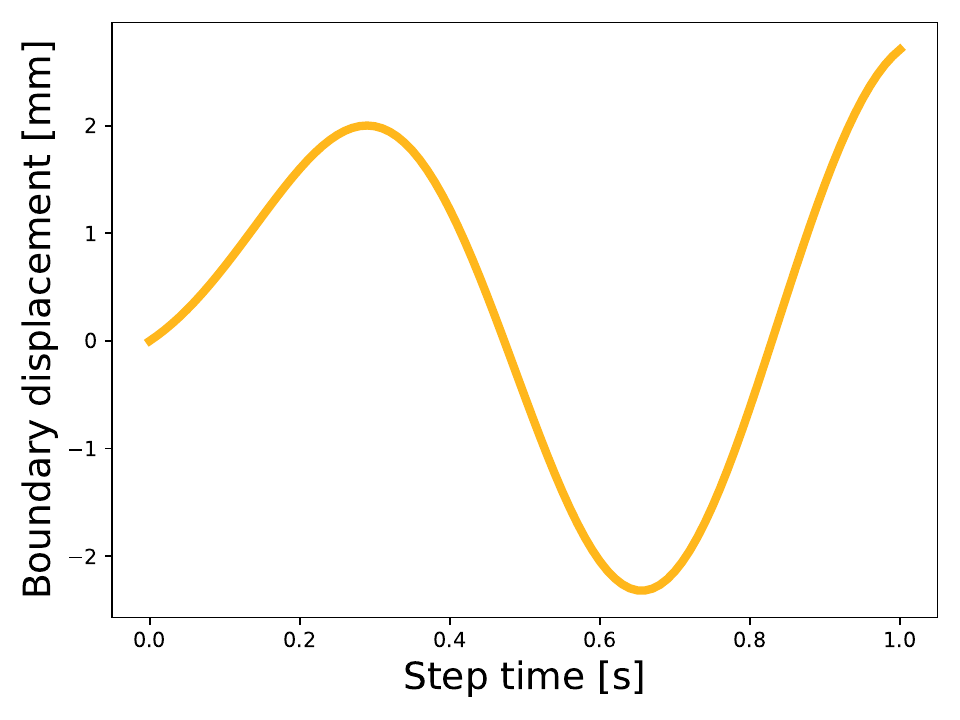}
         \label{disp}
     }
    \caption{\psubref{geom} Schematic of the dog-bone specimen and the mesh used in FE simulations. The axial displacement is along the global X direction. \psubref{disp} A typical applied displacement in the plastic deformation problem.}
    \label{magnitudes}
\end{figure}

In the second example, we consider the plastic deformation of a dog-bone specimen under time-dependent axial loads. The specimen has a length of 110 mm and gauge width of 20 mm. A total of 4756 linear plane stress elements were used, and the specimen is fixed on its left end with displacement applied on the right end. A schematic of the specimen and boundary conditions are shown in \fref{geom}.

In the absence of any body and inertial forces, the equilibrium equations and boundary conditions are:
\begin{equation}
\begin{aligned}
    \nabla \cdot \bm{\sigma} = \bm{0}, \;\; \forall \bm{X} \in \Omega,\\
    \bm{ u } = \Bar{\bm{u}}, \;\; \forall \bm{X} \in \partial \Omega_u,\\
    \label{strong}
\end{aligned}
\end{equation}
where $\bm{\sigma}$ and $\Bar{\bm{u}}$ denote the Cauchy stress and prescribed displacement, respectively. Small-strain assumption is applied, which leads to the following definition of total strain tensor:
\begin{equation}
    \bm{\epsilon} = \frac{1}{2} ( \nabla \bm{u} + \nabla \bm{u}^T ),
    \label{strain}
\end{equation}
as well as its additive decomposition into elastic and plastic strain parts:
\begin{equation}
    \bm{\epsilon} = \bm{\epsilon}^e + \bm{\epsilon}^p.
    \label{decomposition}
\end{equation}
For linear elastic isotropic material in plane stress, stress can be found from elastic strain via:
\begin{equation}
    \begin{bmatrix}
    \sigma_{11} \\
    \sigma_{22} \\
    \sigma_{12}
    \end{bmatrix}
    =
    \frac{E}{1-\nu^2}
    \begin{bmatrix}
    1 & \nu & 0 \\
    \nu & 1 & 0 \\
    0 & 0 & 1-\nu
    \end{bmatrix}
    \begin{bmatrix}
    \epsilon^e_{11} \\
    \epsilon^e_{22} \\
    \epsilon^e_{12}
    \end{bmatrix},
    \label{stress}
\end{equation}
where $E$ and $\nu$ are the Young's modulus and Poisson's ratio. Plasticity is modeled with the simple $J_2$ plasticity with linear isotropic hardening law:
\begin{equation}
    \sigma_y = \sigma_{y0} + H \Bar{\epsilon}_p,
    \label{hardening}
\end{equation}
where $\sigma_y$, $\Bar{\epsilon}_p$, $\sigma_{y0}$, and $H$ denote the flow stress, equivalent plastic strain, initial yield stress, and the hardening modulus, respectively. Relevant material properties are shown in \tref{mat_props}. $\Bar{\epsilon}_p$ is an internal history variable in plasticity that records the accumulation of plastic strain at different points in the loading history. Due to its vital role in plasticity, this is considered one of the output variables of the S-DeepONet, while the other output variable is the von Mises stress:
\begin{equation}
    \Bar{\sigma} = \sqrt{ \sigma_{11}^2 + \sigma_{22}^2 + \sigma_{11}\sigma_{22} + 3\sigma_{12}^2 }.
\end{equation}
\begin{table}[h!]
    \caption{Material properties of the elastic-plastic material model}
    \small
    \centering
    \begin{tabular}{cccccccc}
     Property & \vline & $E$ [MPa] & $\nu$ [/] & $\sigma_{y0}$ [MPa] & H [MPa]\\
    \hline
    Value & \vline  & 2.09$\times 10^{5}$ & 0.3 & 235 & 800\\
    \end{tabular}
    \label{mat_props}
\end{table}

The time-dependent displacement histories were generated similarly to \sref{sec:cfd} with six uniformly spaced control points. The applied displacement is 0 at $t=0$. The displacement magnitude at each control point was randomly selected such that the nominal axial strain magnitude is below 5\%. A typical example of the applied displacement is shown in \fref{disp}. A total of 4000 FE simulations were generated using Abaqus/Standard \cite{Abaqus2021}, and $\Bar{\sigma}$ and $\Bar{\epsilon}_p$ were stored at 40 uniformly spaced time steps as the ground truth labels in the NN training. Similar to \sref{sec:cfd}, some data scaling is needed for best model performance. For this example, the min-max scaler in Scikit-Learn \citep{scikit-learn} was used. Two different scalers were used for the two output components to account for the drastically different scales of stress and plastic strain.

\subsection{Inverse history identification with trained S-DeepONet}
\label{sec:inv}
Once the improved S-DeepONet is trained, it can be used to efficiently infer time-dependent full-field solution contours. Take the example of predicting von Mises stress as described in \sref{sec:plastic_def}. Further post-processing of the predicted fields yields a plot of the mean von Mises stress (over the entire dog bone specimen) as a function of load duration, which is a function of the input load history. If given a known curve of the mean stress over time, it is possible to infer the load history necessary to generate this stress history using the trained S-DeepONet and an optimizer. The outputs of this inverse identification are the five scalar displacement values at the control points, since the smooth load curve in \sref{sec:plastic_def} can be uniquely characterized by these values (the value at first control point is 0). For this work, the genetic algorithm (GA) implementation with PyGAD \cite{gad2021pygad} was used, which provides a gradient-free optimization framework to leverage the trained S-DeepONet as a black-box model. Using GA, 25 generations of optimization were performed with a population size of 100 and the number of parents mating was set to 10. The GA seeks to maximize a scalar fitness function value, and in the current case, it is defined as the inverse of the mean absolute error (MAE) between the predicted and known stress histories. The process of evaluating the fitness value with a trained S-DeepONet is presented in the following algorithm:
\begin{algorithm}[!ht]
\DontPrintSemicolon
    \KwInput{Weights and bias of the trained S-DeepONet, point sets $\bm{P}$ with $N$ points for contour evaluation, five displacement values at control points, desired output stress history $\bm{\sigma}_{ref}$}
    
    \KwOutput{Fitness value}

    \tcc{Initialization}
    Load trained weights and bias to the S-DeepONet model $\mathcal{M}$
    
    \tcc{Get current load history}
    Compute full load curve $\bm{m}$ from control point values with RBI

    \tcc{Forward inference}
    Form a single input set $(\bm{P},\bm{m})$\\
    Infer with $\mathcal{M}$ to obtain output $\bm{O} = \mathcal{M}(\bm{P},\bm{m})$ \\
    Compute predicted mean stress over time $\sigma_{pred,s} = \frac{1}{N} \sum_{i=1}^N O_{is}$\\

    \tcc{Fitness calculation}
    return fitness $F = MAE( \bm{\sigma}_{ref} , \bm{\sigma}_{pred} ) ^{-1}$

\caption{Evaluation of the fitness value}
\label{fitness_algo}
\end{algorithm}

\section{Results and discussion}
\label{sec:results}
All simulations were conducted with six high-end AMD EPYC 7763 Milan CPU cores. All NN training and inference were conducted using a single Nvidia A100 GPU card on Delta, an HPC cluster hosted at the National Center for Supercomputing Applications (NCSA).

To evaluate the model performance in the test set, three quantitative metrics were used, they are the relative $L_2$ error, mean absolute error, and $R^2$ value:
\begin{equation}
\begin{aligned}
    {\rm{Relative \; L_2 \; error}} = \frac{ | f_{FE} - f_{Pred} |_2 }{ |f_{FE}|_2 } \times 100\%,\\
     {\rm{MAE}} = \frac{1}{N_T} \sum_{i=1}^{N_T} \left |  f_{FE} - f_{Pred} \right |,\\
    R^2 = 1 - \frac{ \sum_{i=1}^{N_T} \left( f_{FE} - f_{Pred} \right)^2 }{  \sum_{i=1}^{N_T} \left( f_{FE} - \Bar{f}_{FE} \right)^2 } ,
\end{aligned}
\end{equation}
where $f_{FE}$, $f_{Pred}$, $N_T$, $\Bar{f}_{FE}$ denote the finite element (FE) simulated field value, NN-predicted field value, number of test cases, and mean value of the FE-simulated field values, respectively.

\subsection{Time-dependent fluid flow}
\label{sec:fluid_flow}
To use the improved S-DeepONet to solve the lid-driven cavity problem, we set $C=3$ to predict three vector components $P$, $u$ and $v$ and $S=25$ for 25 output time steps. Doing so yields a model with 800384 trainable parameters. To test the repeatability of the model performance, we trained the S-DeepONet three times with randomly divided training and testing data with a 80-20 data split. The results show similar repeatable performance, with a mean (over all repetitions and components) relative $L_2$ error of 5.891\% and a standard deviation of 0.498\%. Detailed performance metrics for the best performing run out of the three are shown in \tref{metrics_cfd}. The average training and inference (per case) times over the three runs were 22278s and $8\times10^{-3}$s, respectively. While on average, running each FD simulation took 18s, making the inference 2240 times faster than the direct numerical simulation.
\begin{table}[h!]
    \caption{Performance metrics for the CFD example}
    \centering
    \begin{tabular}{ccccc}
     Component & \vline & Relative $L_2$ error [\%] & Mean absolute error & $R^2$ value\\
    \hline
    $P$ & \vline  & 8.244 & 9.671$\times 10^{-3}$ & 0.998 \\

    $u$ & \vline  & 3.465 & 4.060$\times 10^{-3}$ & 0.999 \\

    $v$ & \vline  & 6.717 & 1.198$\times 10^{-3}$ & 0.998 \\
    
    \end{tabular}
    \label{metrics_cfd}
\end{table}

\begin{figure}[h!]
\newcommand\x{0.23}
    \centering
    \begin{tabular}{ c c c c c }
    \begin{minipage}[c]{\x\textwidth}
       \centering 
        \subfloat[Step 1, best]{\includegraphics[trim={.9cm 1.cm .9cm 1.6cm},clip,width=\textwidth]{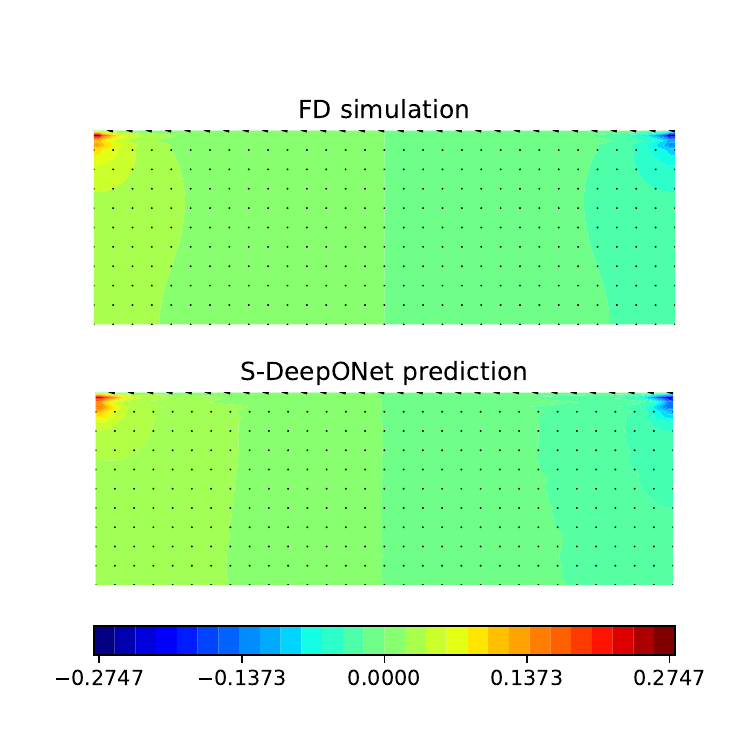}
        \label{p11}}
    \end{minipage} &
    \begin{minipage}[c]{\x\textwidth}
       \centering 
        \subfloat[Step 9, best]{\includegraphics[trim={.9cm 1.cm .9cm 1.6cm},clip,width=\textwidth]{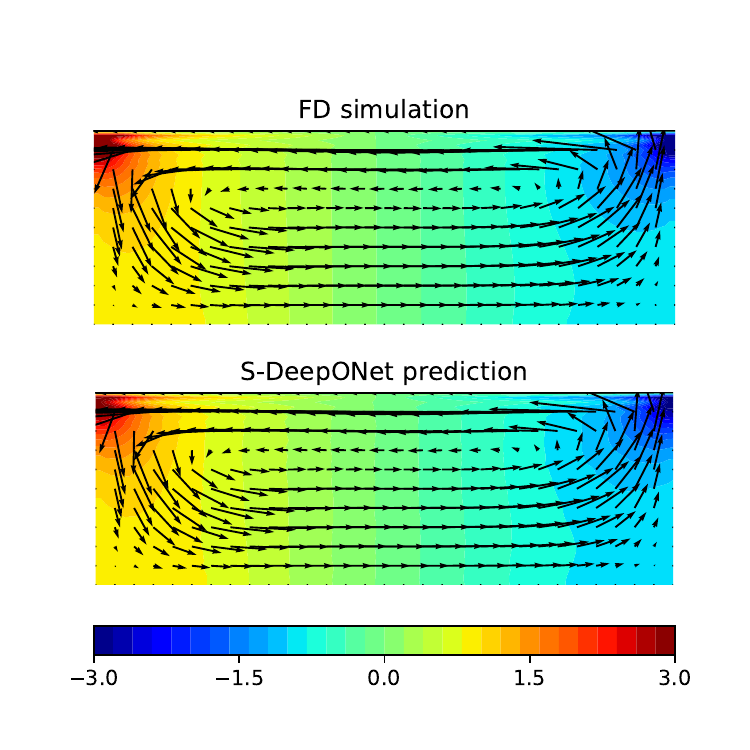}
        \label{p12}}
    \end{minipage} &
    \begin{minipage}[c]{\x\textwidth}
       \centering 
        \subfloat[Step 17, best]{\includegraphics[trim={.9cm 1.cm .9cm 1.6cm},clip,width=\textwidth]{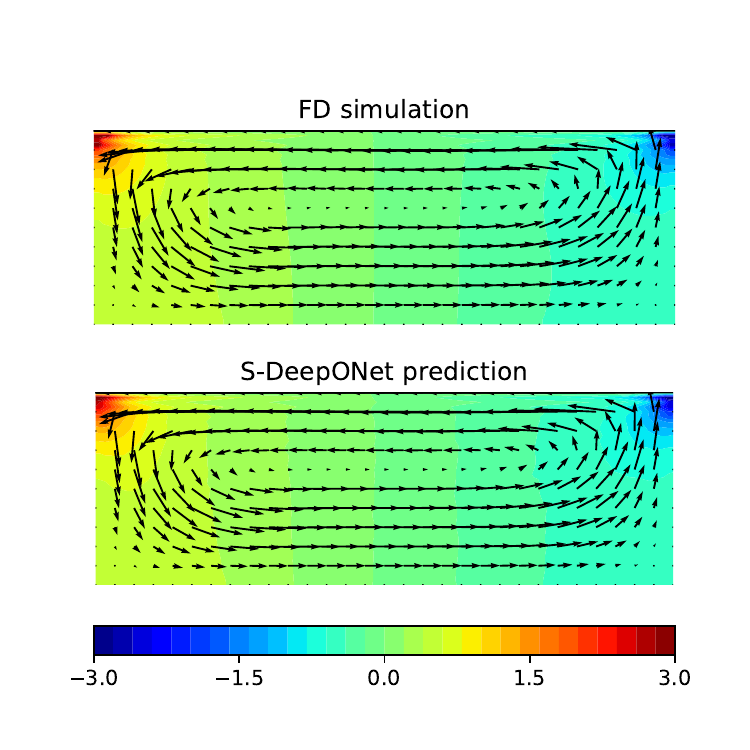}
        \label{p13}}
    \end{minipage} &
    \begin{minipage}[c]{\x\textwidth}
       \centering 
        \subfloat[Step 25, best]{\includegraphics[trim={.9cm 1.cm .9cm 1.6cm},clip,width=\textwidth]{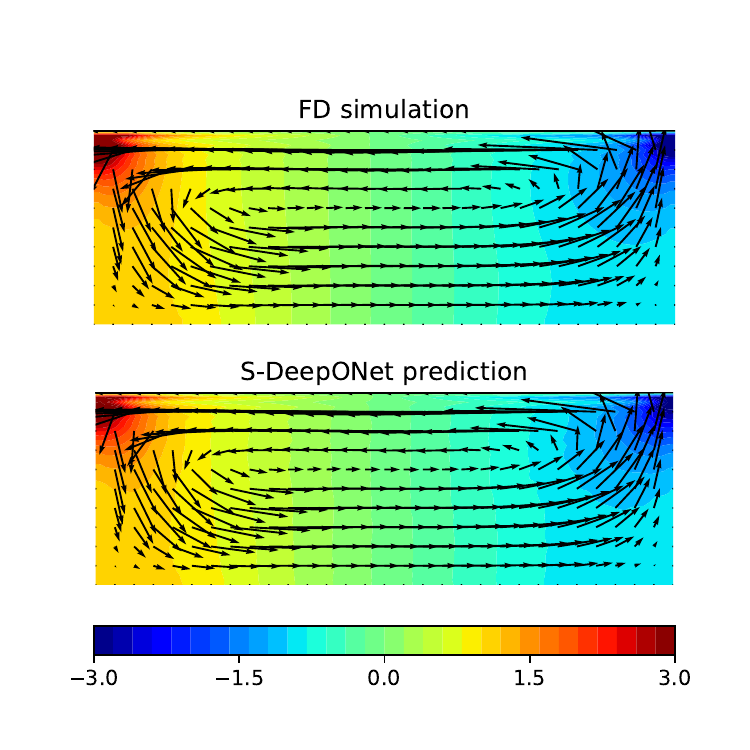}
        \label{p14}}
    \end{minipage} \\

    \begin{minipage}[c]{\x\textwidth}
       \centering 
        \subfloat[Step 1, 80$^{th}$ pct]{\includegraphics[trim={.9cm 1.cm .9cm 1.6cm},clip,width=\textwidth]{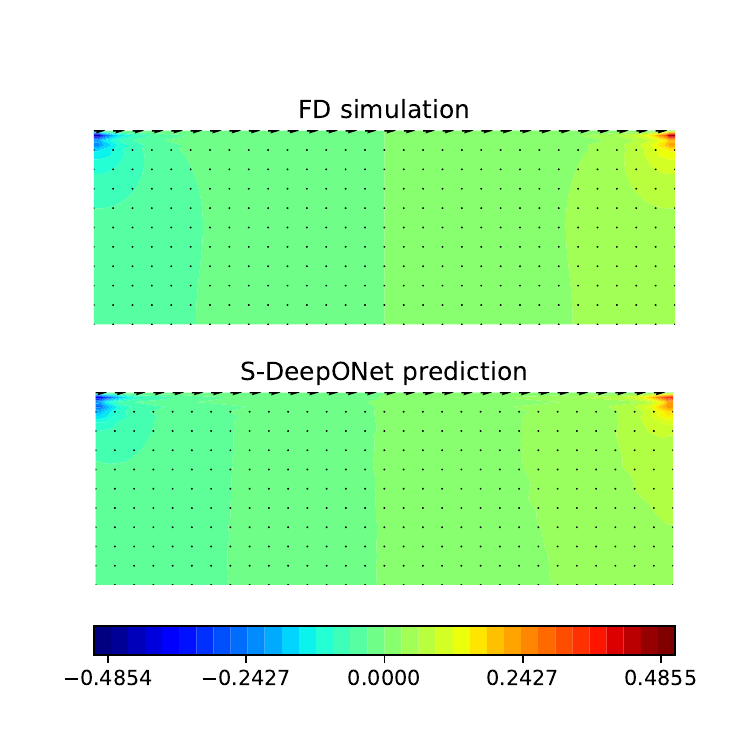}
        \label{p21}}
    \end{minipage} &
    \begin{minipage}[c]{\x\textwidth}
       \centering 
        \subfloat[Step 9, 80$^{th}$ pct]{\includegraphics[trim={.9cm 1.cm .9cm 1.6cm},clip,width=\textwidth]{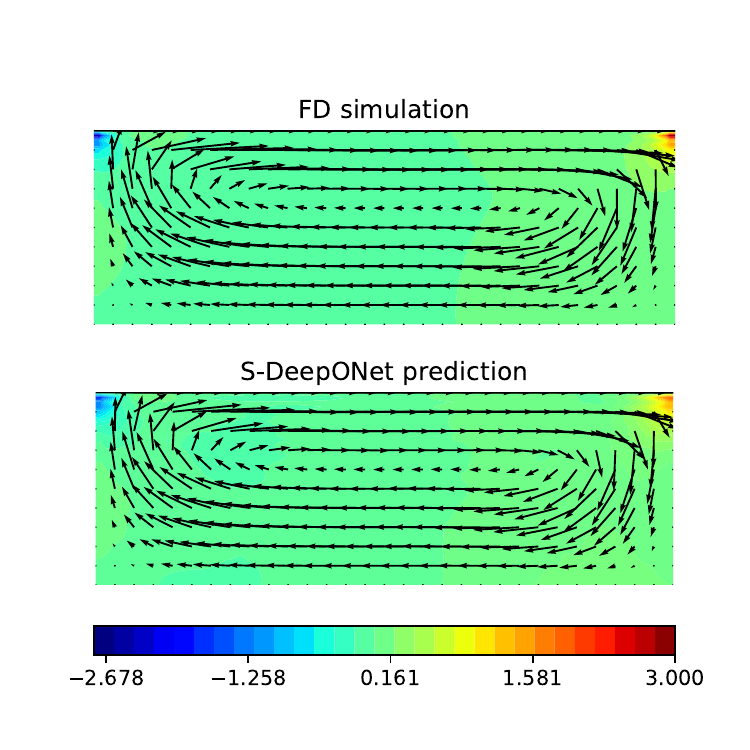}
        \label{p22}}
    \end{minipage} &
    \begin{minipage}[c]{\x\textwidth}
       \centering 
        \subfloat[Step 17, 80$^{th}$ pct]{\includegraphics[trim={.9cm 1.cm .9cm 1.6cm},clip,width=\textwidth]{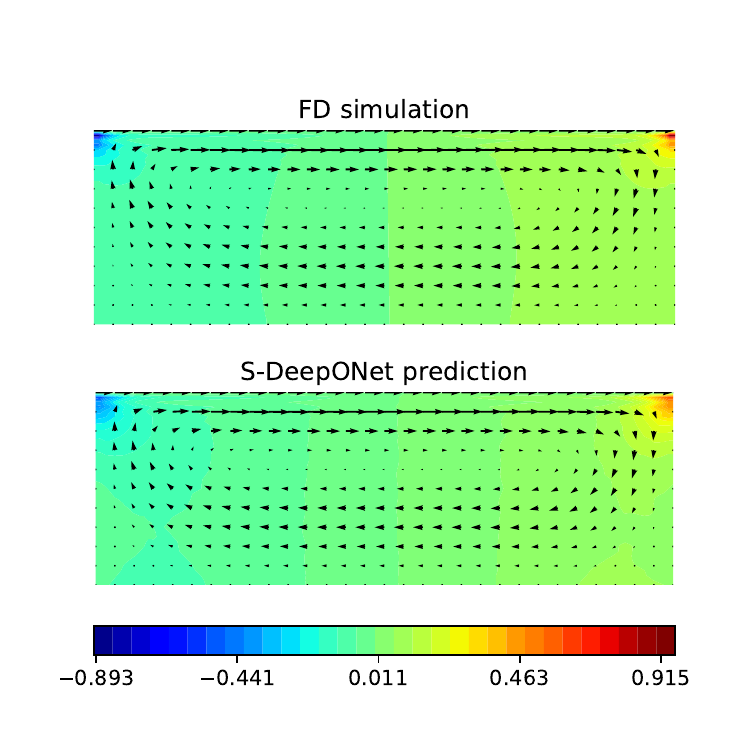}
        \label{p23}}
    \end{minipage} &
    \begin{minipage}[c]{\x\textwidth}
       \centering 
        \subfloat[Step 25, 80$^{th}$ pct]{\includegraphics[trim={.9cm 1.cm .9cm 1.6cm},clip,width=\textwidth]{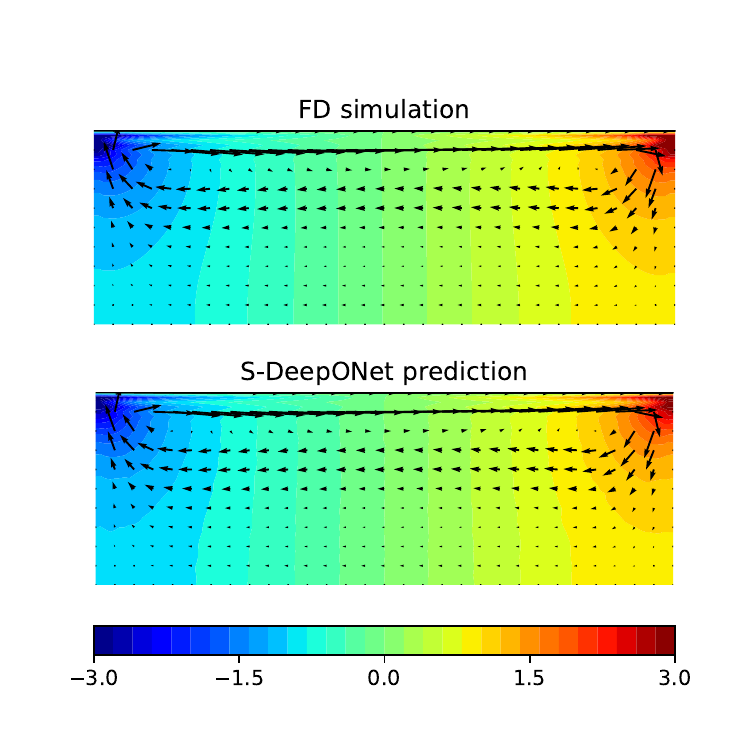}
        \label{p24}}
    \end{minipage} \\

    \begin{minipage}[c]{\x\textwidth}
       \centering 
        \subfloat[Step 1, worst]{\includegraphics[trim={.9cm 1.cm .9cm 1.6cm},clip,width=\textwidth]{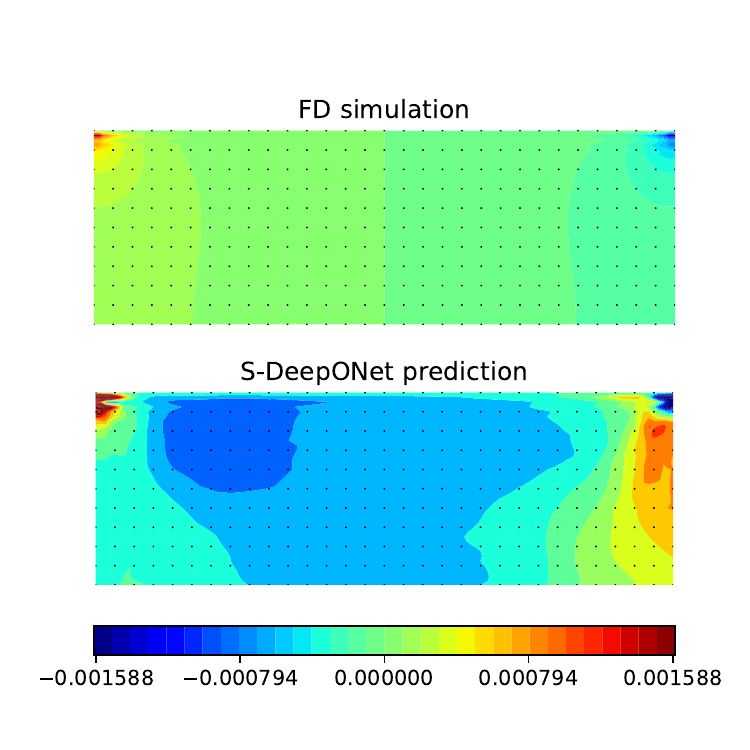}
        \label{p31}}
    \end{minipage} &
    \begin{minipage}[c]{\x\textwidth}
       \centering 
        \subfloat[Step 9, worst]{\includegraphics[trim={.9cm 1.cm .9cm 1.6cm},clip,width=\textwidth]{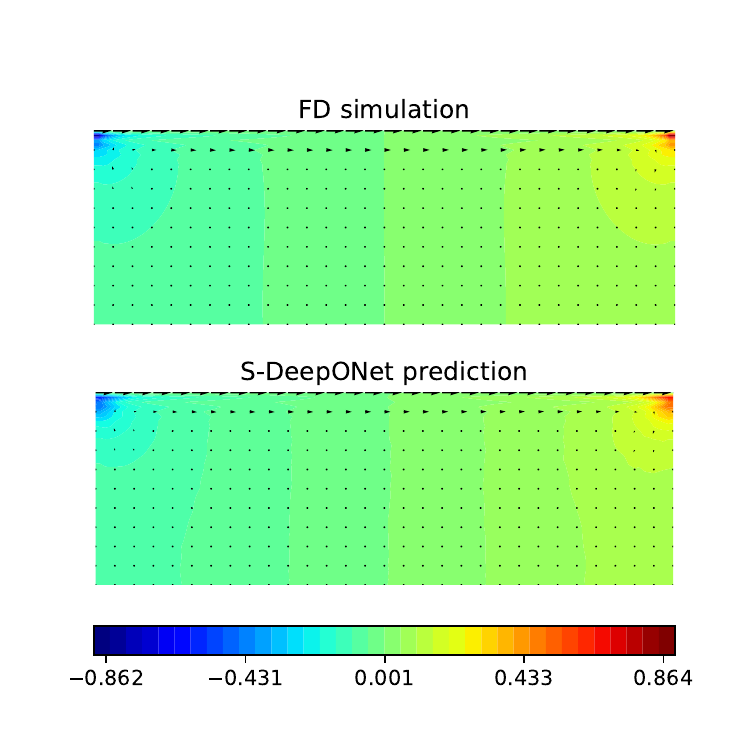}
        \label{p32}}
    \end{minipage} &
    \begin{minipage}[c]{\x\textwidth}
       \centering 
        \subfloat[Step 17, worst]{\includegraphics[trim={.9cm 1.cm .9cm 1.6cm},clip,width=\textwidth]{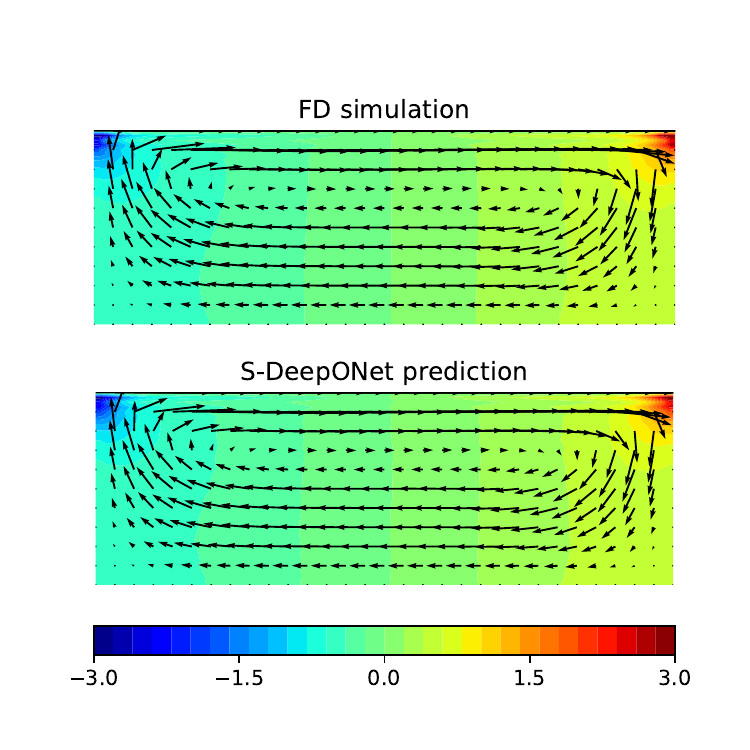}
        \label{p33}}
    \end{minipage} &
    \begin{minipage}[c]{\x\textwidth}
       \centering 
        \subfloat[Step 25, worst]{\includegraphics[trim={.9cm 1.cm .9cm 1.6cm},clip,width=\textwidth]{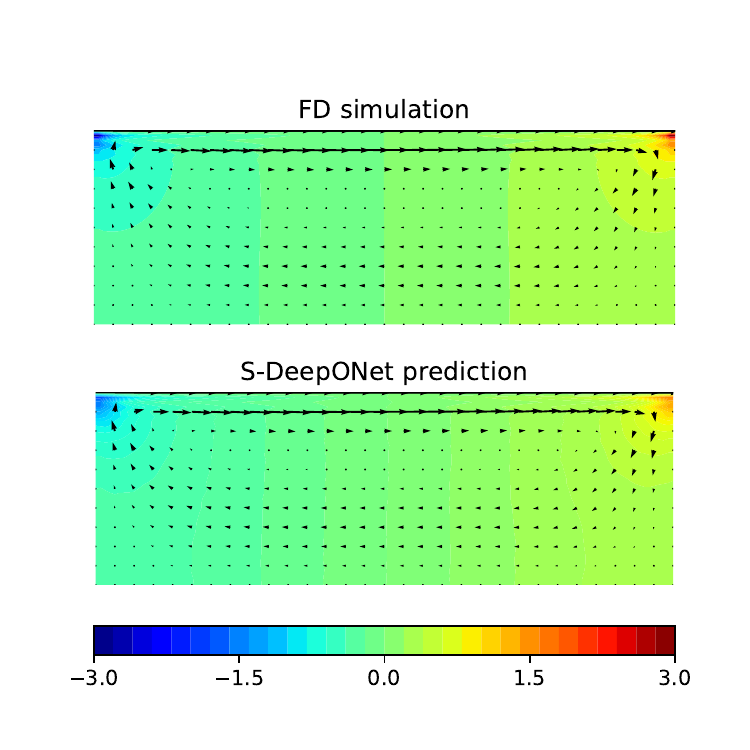}
        \label{p34}}
    \end{minipage} \\
    
    \end{tabular}
    \caption{Contour and quiver plots for the flow field at different time and percentiles of prediction accuracy. In each sub-figure, the top and bottom rows show finite difference solution and S-DeepONet prediction, resp. The background contour is colored by the pressure, and the flow velocity field is rendered as arrows with length proportional to the velocity magnitude.}
    \label{pct_plot2}
\end{figure}

The contour and quiver plots for pressure and velocity vector at different time steps are shown in \fref{pct_plot2}. They are ranked by the percentile of mean relative $L_2$ error in the predictions, and the 0$^{th}$ (best case), 80$^{th}$ and 100$^{th}$ (worst case) percentiles are shown for representation. For all plots in \fref{pct_plot2}, the filled contour was colored by pressure, and the velocity vectors are shown as arrows in the domain whose lengths are proportional to the velocity magnitude. 

\begin{figure}[h!] 
    \centering
     \subfloat[]{
         \includegraphics[trim={0cm 0cm 0cm .2cm},clip,width=0.3\textwidth]{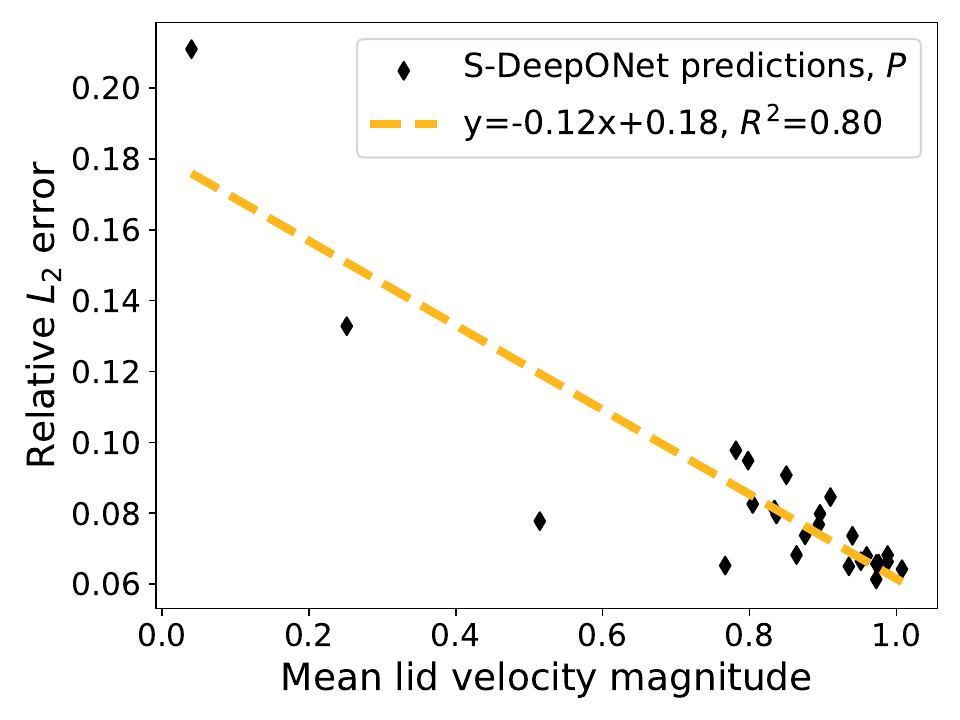}
         \label{lp3}
     }
     \subfloat[]{
         \includegraphics[trim={0cm 0cm 0cm .2cm},clip,width=0.3\textwidth]{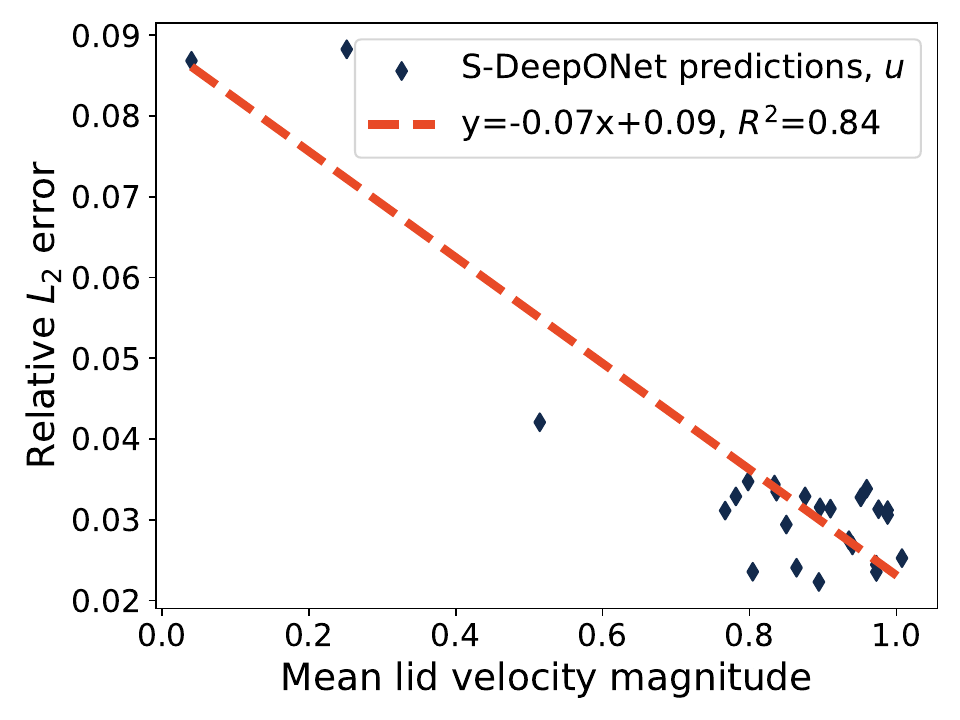}
         \label{lp4}
     }
     \subfloat[]{
         \includegraphics[trim={0cm 0cm 0cm .2cm},clip,width=0.3\textwidth]{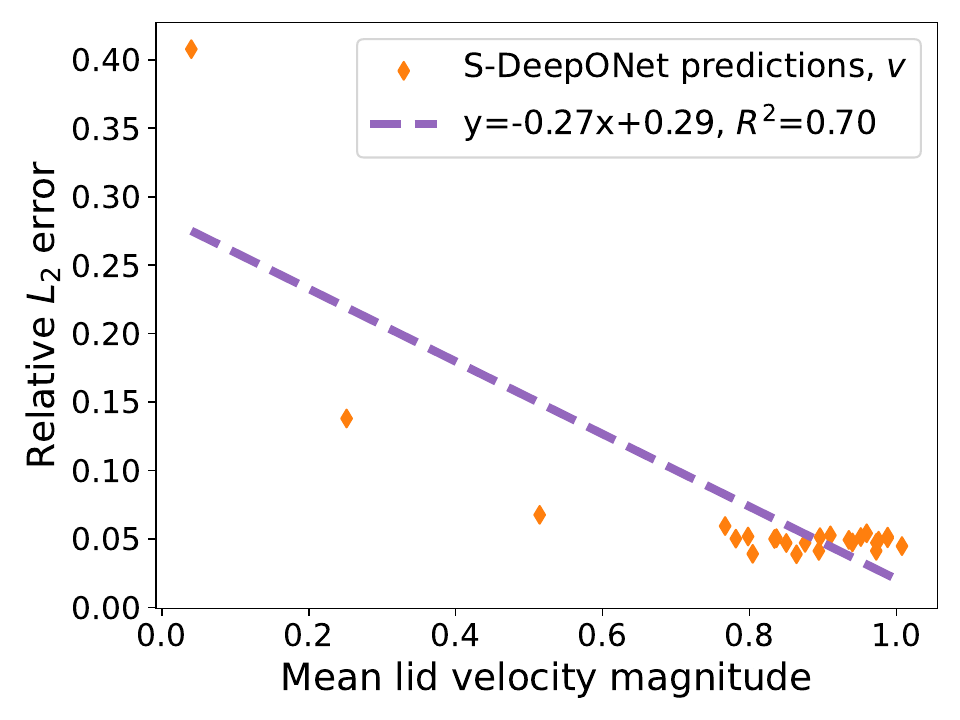}
         \label{lp5}
     }
    \caption{Scatter plots and trend lines of case-averaged errors for: \psubref{lp3} Pressure ($P$), \psubref{lp4} x velocity ($u$), \psubref{lp5} y velocity ($v$).}
    \label{err_vs_u_cfd}
\end{figure}

When predicting full-field solution fields at multiple time steps, two types of errors are worth investigating. They are the time-averaged error (i.e., for each prediction case, the average error over all prediction time steps) and case-averaged error (i.e., for each time step, the average error over all prediction cases). Time-averaged error provides a holistic measurement of model prediction accuracy over all time steps, while the case-averaged error provides insight on how the prediction accuracy changes with the magnitude of the input function. For that, scatter plots of the relative $L_2$ prediction versus the mean magnitude of the lid velocity are shown in \fref{err_vs_u_cfd}.

From the performance metrics in \tref{metrics_cfd}, we see that the S-DeepONet is able to generate accurate predictions with relative $L_2$ errors less than 10\% and $R^2$ of above 0.99 for all components with only 3200 training data point. The contour plots in \fref{pct_plot2} further confirm the accuracy of the model, as we can see that the S-DeepONet predicted contours match closely with those computed from direct numerical simulations. For the worst case scenario (last row of \fref{pct_plot2}), we see the pressure contour prediction is inaccurate at step 1, where the pressure magnitude is small. Even in this worst case (ranked by time-averaged error), we see that the model is able to generate accurate predictions for some time steps (e.g., step 17, third column), where the magnitudes of the solution fields are reasonable. The relationships between prediction error and mean lid velocity magnitude is further elucidated by \fref{err_vs_u_cfd}, where we see a consistent decreasing trend for increasing lid velocity magnitude. All three components in this case show relatively strong correlations with the trend line $R^2 \ge 0.7$, indicating that the prediction error has a strong correlation with the lid velocity, a subject we will further investigate and compare in \sref{sec:plastic_res}.

\subsection{History-dependent plastic deformation}
\label{sec:plastic_res}
\begin{figure}[h!]
\newcommand\x{0.32}
    \centering
    \begin{tabular}{ c c c c c }
    \begin{minipage}[c]{\x\textwidth}
       \centering 
        \subfloat[Step 1, best]{\includegraphics[trim={.5cm .5cm .5cm 1.cm},clip,width=\textwidth]{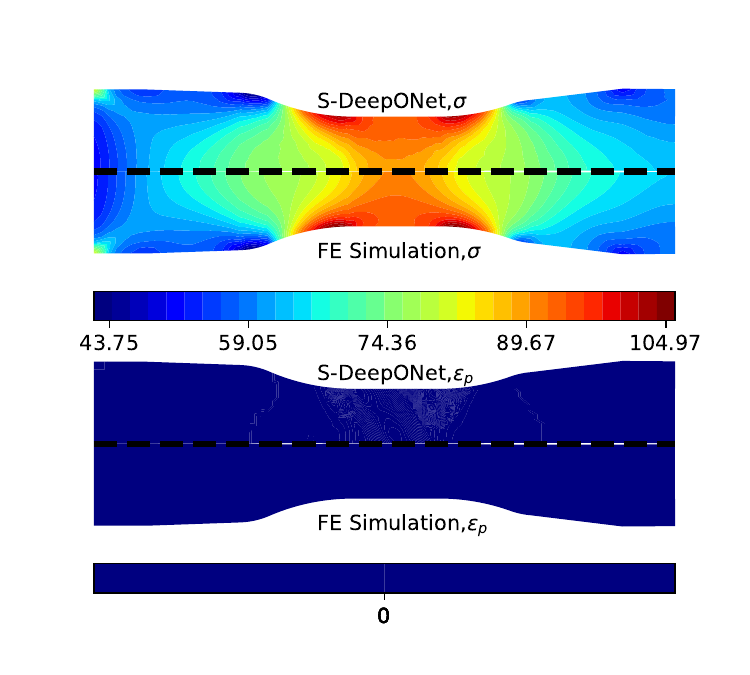}
        \label{p11}}
    \end{minipage} &
    \begin{minipage}[c]{\x\textwidth}
       \centering 
        \subfloat[Step 20, best]{\includegraphics[trim={.5cm .5cm .5cm 1.cm},clip,width=\textwidth]{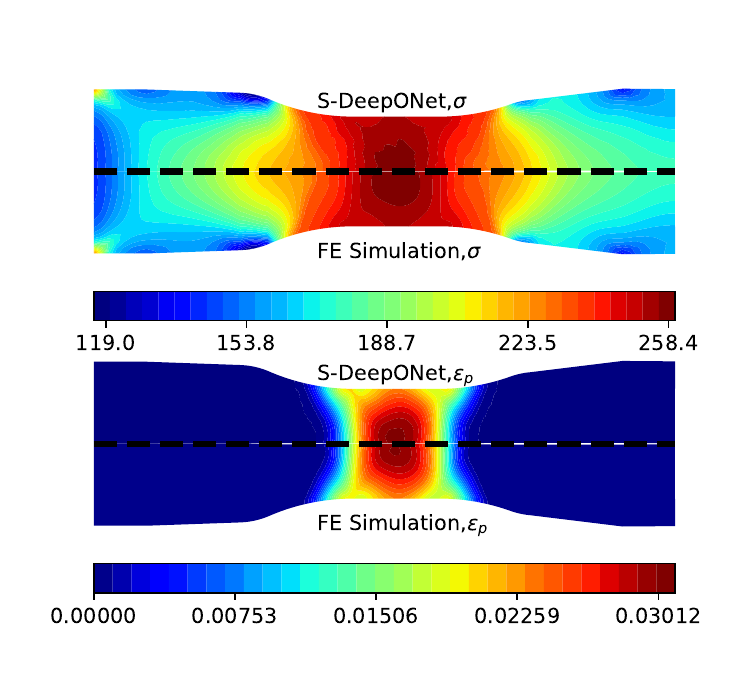}
        \label{p12}}
    \end{minipage} &
    \begin{minipage}[c]{\x\textwidth}
       \centering 
        \subfloat[Step 40, best]{\includegraphics[trim={.5cm .5cm .5cm 1.cm},clip,width=\textwidth]{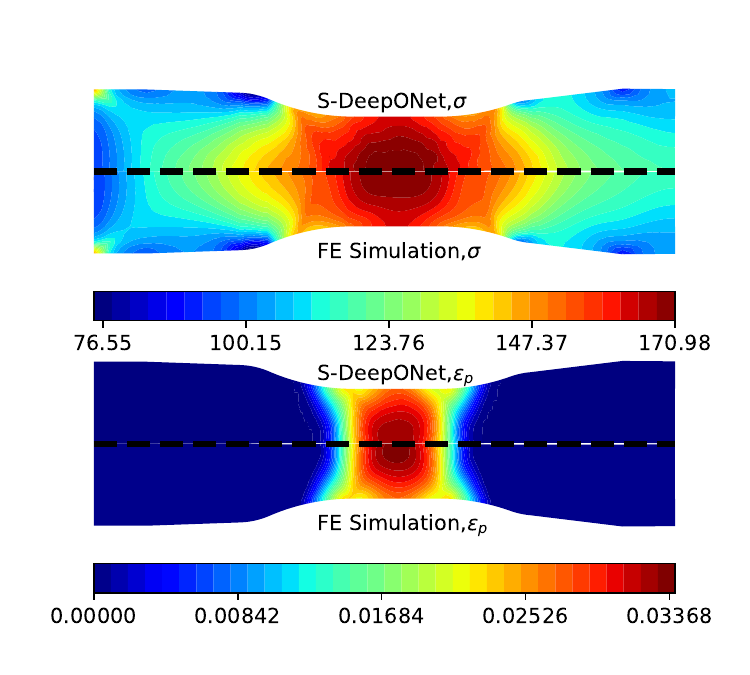}
        \label{p13}}
    \end{minipage} \\

    \begin{minipage}[c]{\x\textwidth}
       \centering 
        \subfloat[Step 1, 80$^{th}$ pct]{\includegraphics[trim={.5cm .5cm .5cm 1.cm},clip,width=\textwidth]{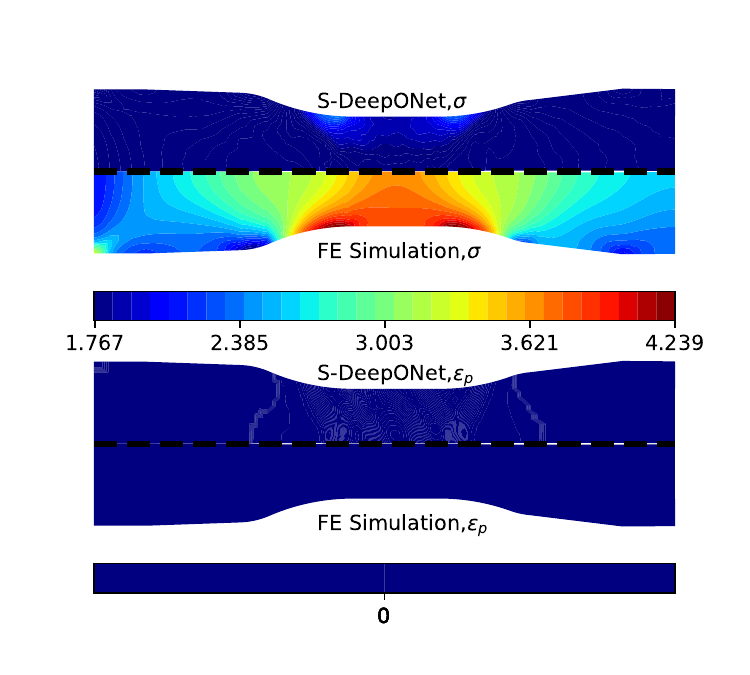}
        \label{p21}}
    \end{minipage} &
    \begin{minipage}[c]{\x\textwidth}
       \centering 
        \subfloat[Step 20, 80$^{th}$ pct]{\includegraphics[trim={.5cm .5cm .5cm 1.cm},clip,width=\textwidth]{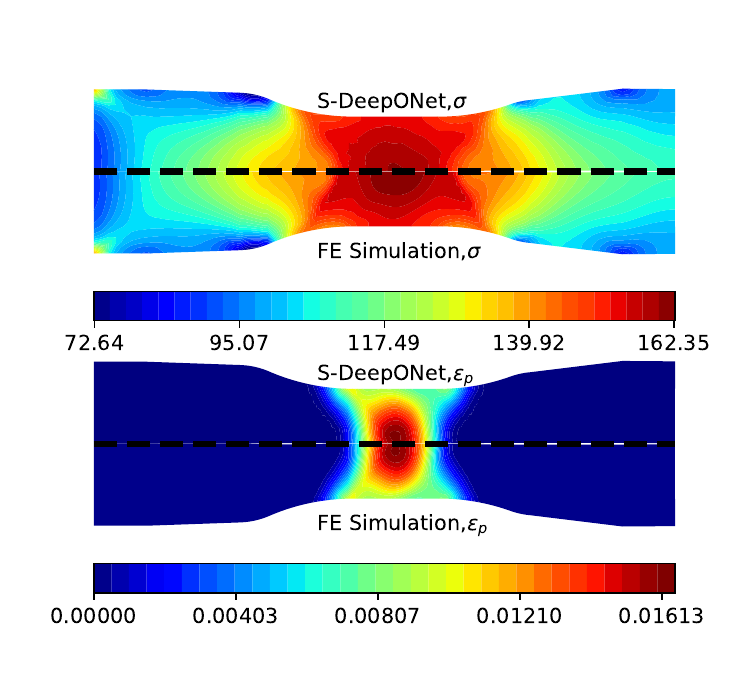}
        \label{p22}}
    \end{minipage} &
    \begin{minipage}[c]{\x\textwidth}
       \centering 
        \subfloat[Step 40, 80$^{th}$ pct]{\includegraphics[trim={.5cm .5cm .5cm 1.cm},clip,width=\textwidth]{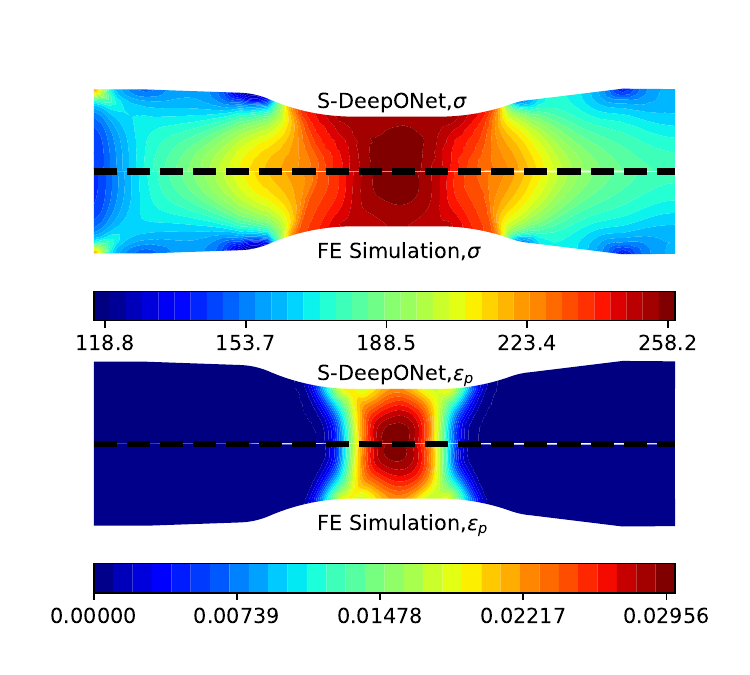}
        \label{p23}}
    \end{minipage} \\

    \begin{minipage}[c]{\x\textwidth}
       \centering 
        \subfloat[Step 1, worst]{\includegraphics[trim={.5cm .5cm .5cm 1.cm},clip,width=\textwidth]{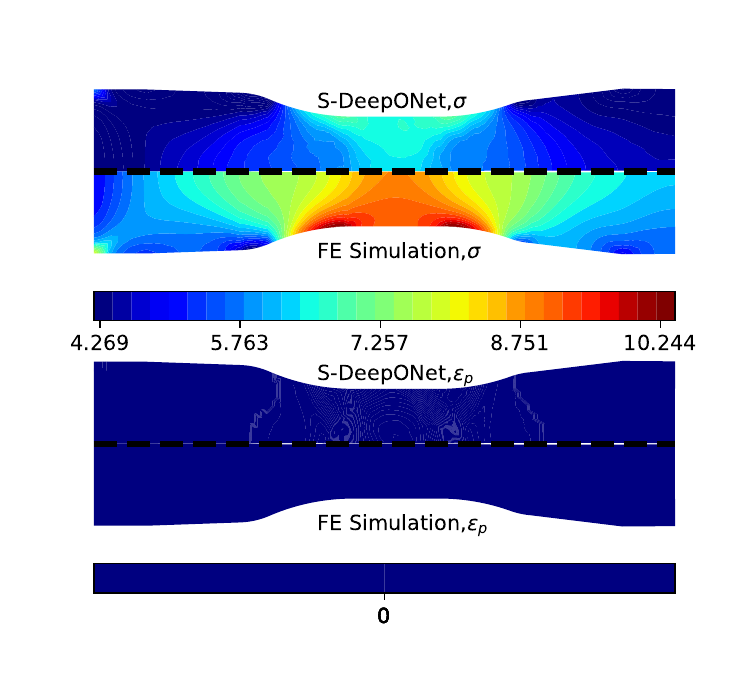}
        \label{p31}}
    \end{minipage} &
    \begin{minipage}[c]{\x\textwidth}
       \centering 
        \subfloat[Step 20, worst]{\includegraphics[trim={.5cm .5cm .5cm 1.cm},clip,width=\textwidth]{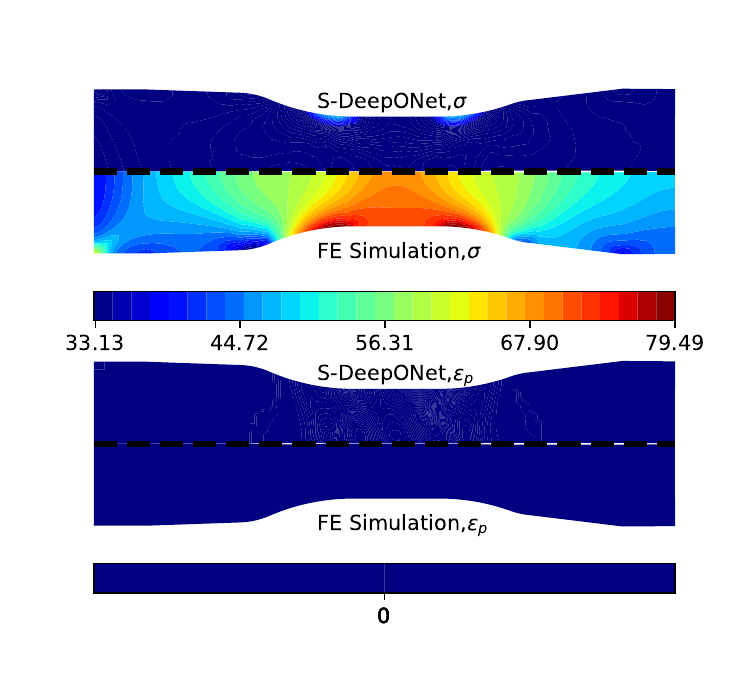}
        \label{p32}}
    \end{minipage} &
    \begin{minipage}[c]{\x\textwidth}
       \centering 
        \subfloat[Step 40, worst]{\includegraphics[trim={.5cm .5cm .5cm 1.cm},clip,width=\textwidth]{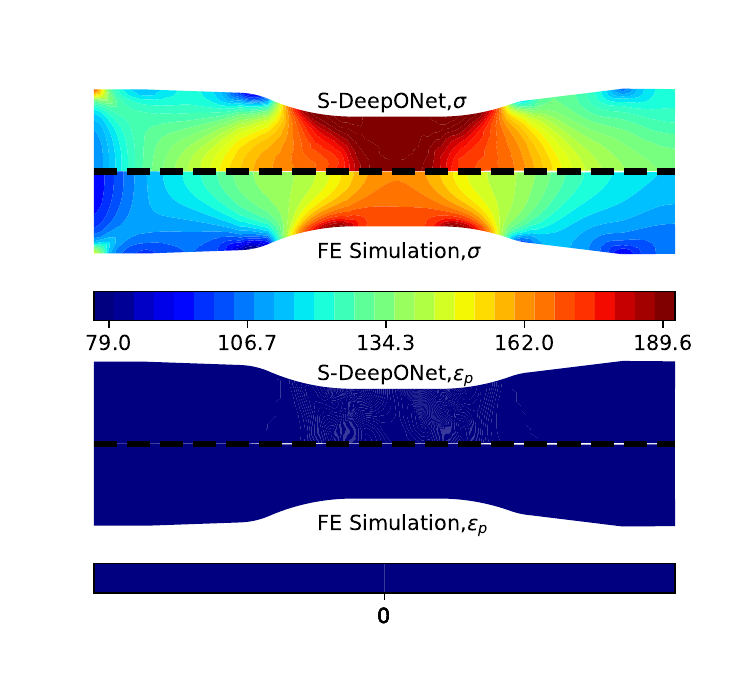}
        \label{p33}}
    \end{minipage} \\
    
    \end{tabular}
    \caption{Contour plots for stress and plastic strain fields at different time and percentiles of prediction accuracy. In each sub-figure, the top and bottom rows show stress and plastic strain, resp. In each contour, only top half of NN-prediction is shown due to symmetry, with FE solution shown in bottom half.}
    \label{pct_plot1}
\end{figure}

The means and standard deviations of the metrics over three repetitions of the S-DeepONet training are summarized in \tref{cv1}.
\begin{table}[h!]
    \caption{Result repeatability, plastic deformation}
    \centering
    \begin{tabular}{ccccc}
     Component & \vline & Relative $L_2$ error [\%] & Mean absolute error & $R^2$ value\\
    \hline
    $\Bar{\sigma}$ & \vline  & 5.164 (8.691$\times 10^{-2}$) \tablefootnote{Data format: Mean (Standard deviation)} & 2.544 (6.039$\times 10^{-2}$) MPa & 0.997 (9.764$\times 10^{-5}$) \\

    $\Bar{\epsilon}_p$ & \vline  & 21.190 (0.190) & 8.376$\times 10^{-5}$ (9.084$\times 10^{-7}$) & 0.999 (1.279$\times 10^{-4}$) \\
    
    \end{tabular}
    \label{cv1}
\end{table}
To show the statistical distribution of prediction error among the 800 test cases, stress contours that correspond to the 0$^{th}$ (best case), 80$^{th}$, and 100$^{th}$ (worst case) percentile prediction error are displayed in \fref{pct_plot1} for the improved S-DeepONet model. For the ranking of prediction errors, the average relative $L_2$ error over all time steps and all output components were used to obtain a single scalar representation of prediction accuracy over all time steps and output components. Due to the symmetry in the loading, only the top half of the NN-predictions are shown in \fref{pct_plot1}, with the bottom half reserved for (the flipped top half of) the FE simulation, the two are separated by a black dashed line.

\begin{figure}[h!]
\newcommand\x{0.23}
    \centering
    \begin{tabular}{ c c c c c }

    \begin{minipage}[c]{\x\textwidth}
       \centering 
        \subfloat[Step 1, 50$^{th}$ pct]{\includegraphics[trim={.5cm .5cm .5cm 1.cm},clip,width=\textwidth]{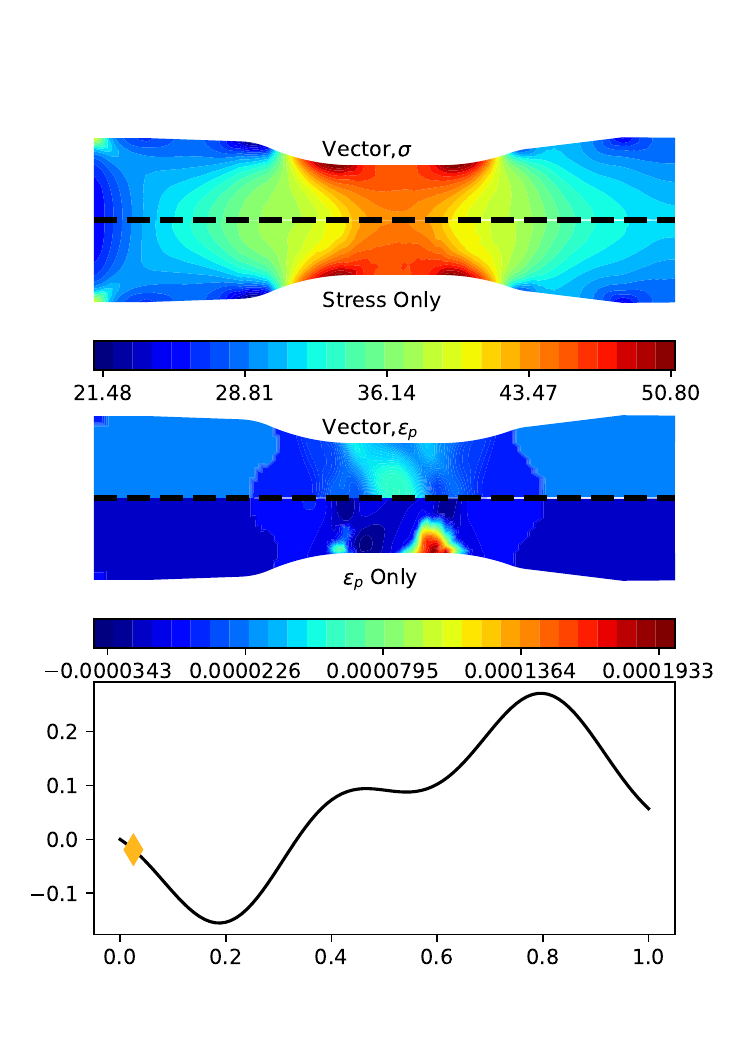}
        \label{p41}}
    \end{minipage} &
    \begin{minipage}[c]{\x\textwidth}
       \centering 
        \subfloat[Step 14, 50$^{th}$ pct]{\includegraphics[trim={.5cm .5cm .5cm 1.cm},clip,width=\textwidth]{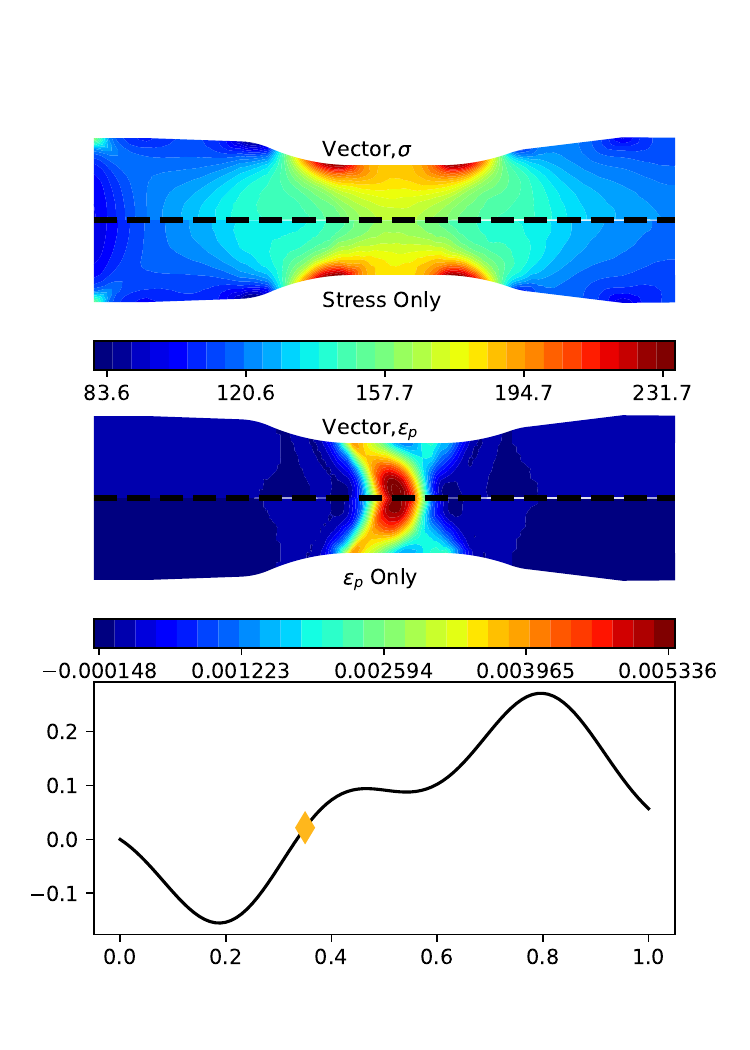}
        \label{p42}}
    \end{minipage} &
    \begin{minipage}[c]{\x\textwidth}
       \centering 
        \subfloat[Step 27, 50$^{th}$ pct]{\includegraphics[trim={.5cm .5cm .5cm 1.cm},clip,width=\textwidth]{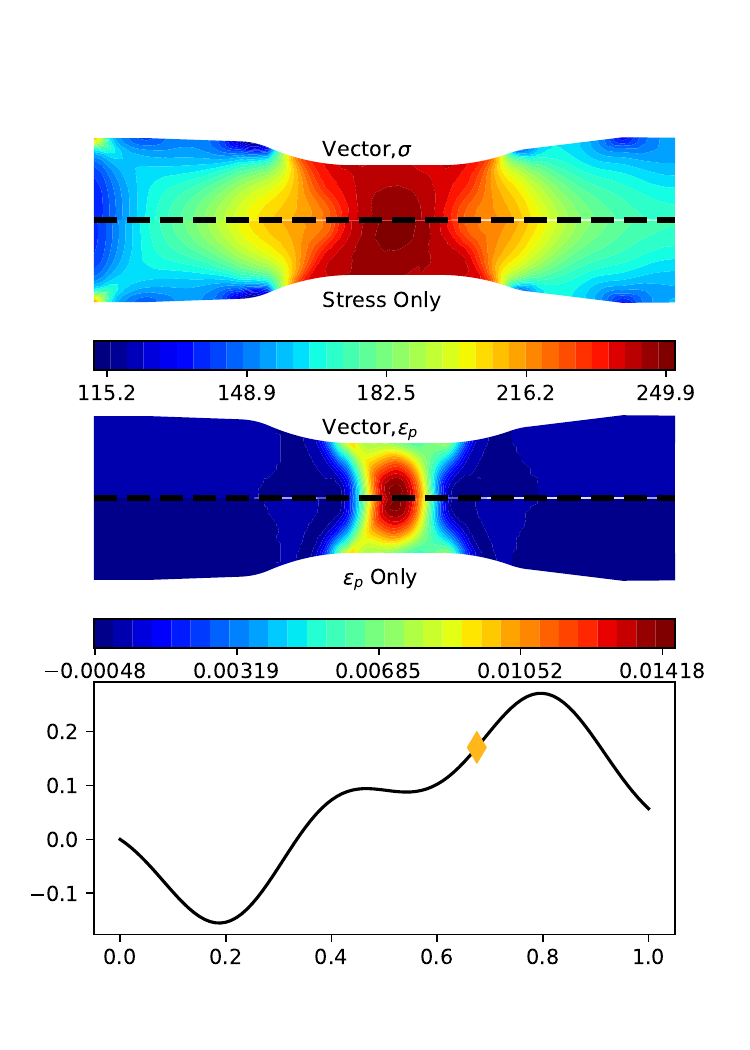}
        \label{p43}}
    \end{minipage} &
    \begin{minipage}[c]{\x\textwidth}
       \centering 
        \subfloat[Step 40, 50$^{th}$ pct]{\includegraphics[trim={.5cm .5cm .5cm 1.cm},clip,width=\textwidth]{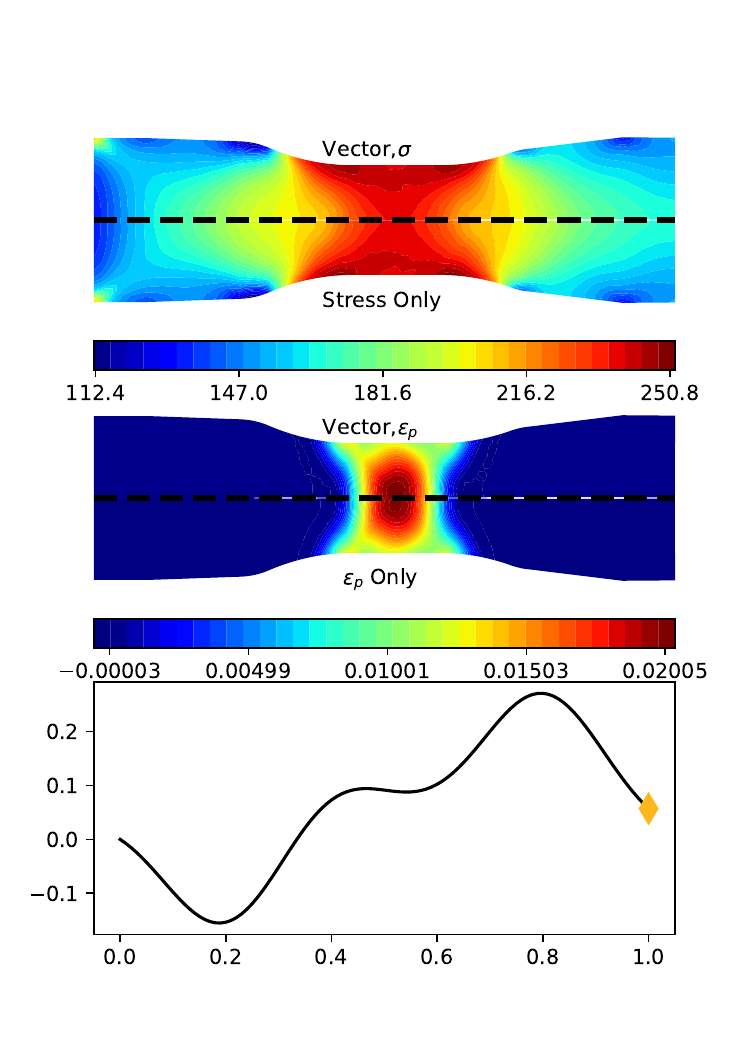}
        \label{p44}}
    \end{minipage} \\
    
    \end{tabular}
    \caption{Contour plots for stress and plastic strain fields predicted by the vector and scalar S-DeepONets. In each sub-figure, the top and bottom rows show stress and plastic strain, resp. In each contour, only top half of the contour is shown due to symmetry. The top half shows the prediction from the vector S-DeepONet, and the bottom half shows that from the scalar S-DeepONets.}
    \label{model_comp}
\end{figure}

To highlight the efficiency of the current vector-output model, we trained two scalar S-DeepONets (i.e., by setting $C=1$), one for $\Bar{\sigma}$ and the other for $\Bar{\epsilon}_p$, and compare the computational efficiency and accuracy of the models. For a fair comparison, training of the three models were conducted using identical training and testing data points. The model size, training, and inference time are compared in \tref{comp1}.
\begin{table}[h!]
    \caption{Vector model vs. scalar models}
    \centering
    \begin{tabular}{ccccc}
     Model & \vline & Trainable parameters & Training time [s] & Inference time [s] \\
    \hline
    Vector & \vline  & 800975 & 15547 & 4.025$\times 10^{-3}$ \\
    Scalar, $\Bar{\sigma}$ & \vline  & 797711 & 9424 & 3.675$\times 10^{-3}$ \\
    Scalar, $\Bar{\epsilon}_p$ & \vline  & 797711 & 9364 & 3.588$\times 10^{-3}$ \\    
    \end{tabular}
    \label{comp1}
\end{table}
The prediction quality metrics are summarized in \tref{comp2}.
\begin{table}[h!]
    \caption{Vector model vs. scalar models, performance metrics}
    \centering
    \begin{tabular}{cccccccc}
     Model & \vline & Rel. $L_2$,$\Bar{\sigma}$  & Mean abs.,$\Bar{\sigma}$  & $R^2$,$\Bar{\sigma}$ & Rel. $L_2$,$\Bar{\epsilon}_p$  & Mean abs.,$\Bar{\epsilon}_p$  & $R^2$,$\Bar{\epsilon}_p$\\
    \hline
    Vector & \vline  & 5.164 & 2.544 & 0.997 & 21.190 & 8.376$\times 10^{-5}$ & 0.999 \\
    Scalar, $\Bar{\sigma}$ & \vline  & 5.185 & 2.648 & 0.996 & / & / & / \\
    Scalar, $\Bar{\epsilon}_p$ & \vline  & / & / & / & 19.471 & 5.549$\times 10^{-5}$ & 0.999 \\    
    \end{tabular}
    \label{comp2}
\end{table}
The contour plots predicted by the vector S-DeepONet model and the two scalar S-DeepONet models at different time steps are shown in \fref{model_comp} for model comparison. The test case corresponding to the 50$^{th}$ (median case) percentile prediction error is shown for representation.

\begin{figure}[h!] 
    \centering
     \subfloat[]{
         \includegraphics[trim={0cm 0cm 0cm .2cm},clip,width=0.45\textwidth]{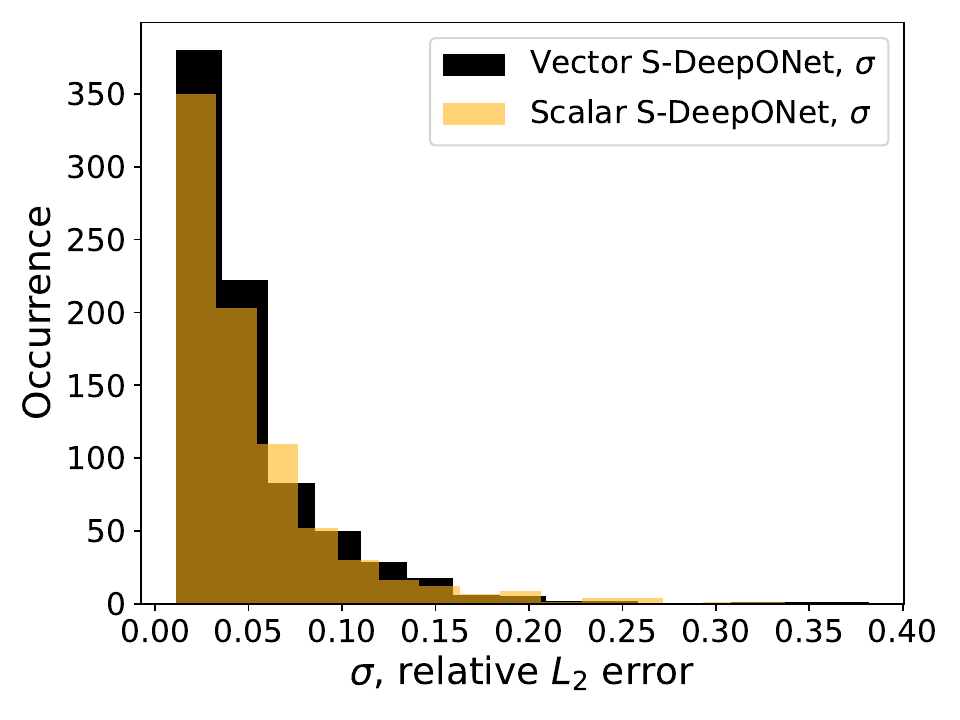}
         \label{htr1}
     }
     \subfloat[]{
         \includegraphics[trim={0cm 0cm 0cm .2cm},clip,width=0.45\textwidth]{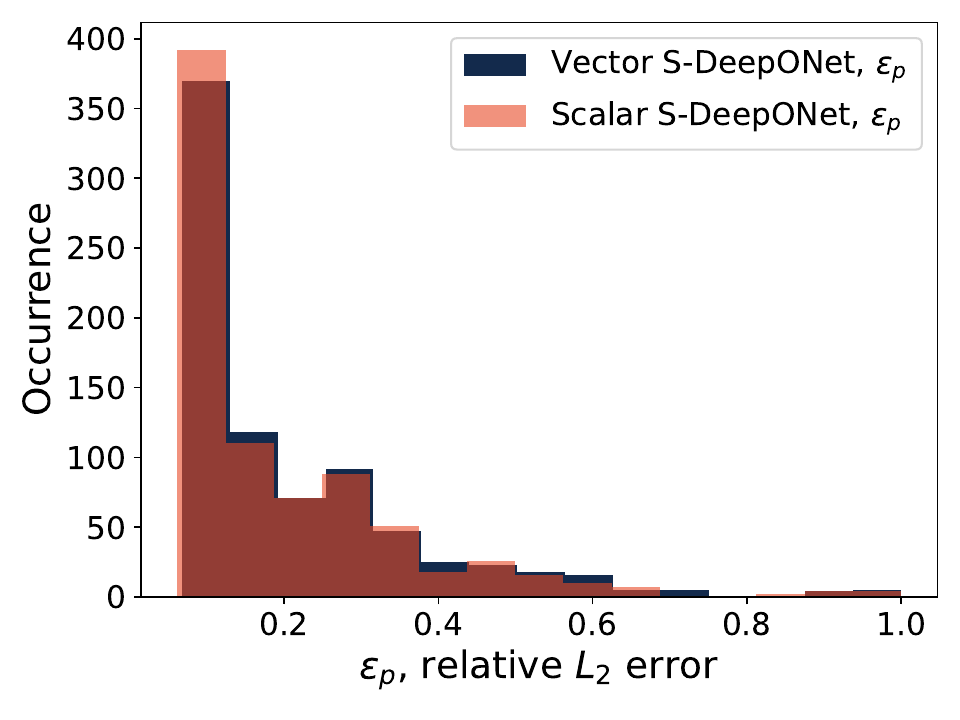}
         \label{htr2}
     }
    \caption{Histograms of time-averaged errors for: \psubref{htr1} von Mises stress, \psubref{htr2} equivalent plastic strain.}
    \label{model_comp_hist}
\end{figure}

\begin{figure}[h!] 
    \centering
     \subfloat[]{
         \includegraphics[trim={0cm 0cm 0cm .2cm},clip,width=0.45\textwidth]{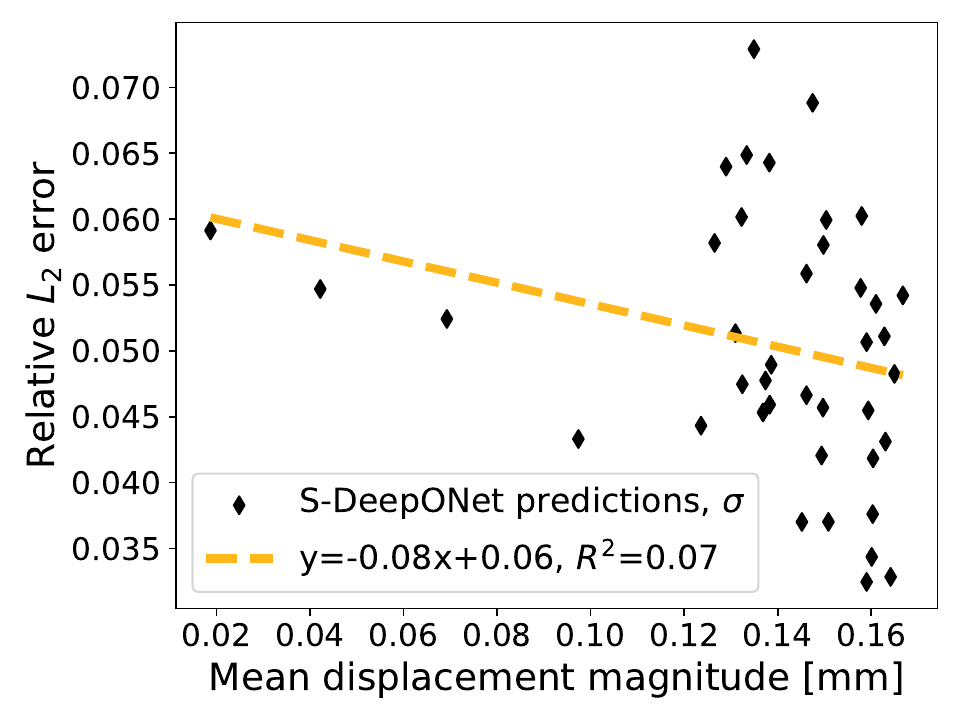}
         \label{lp1}
     }
     \subfloat[]{
         \includegraphics[trim={0cm 0cm 0cm .2cm},clip,width=0.45\textwidth]{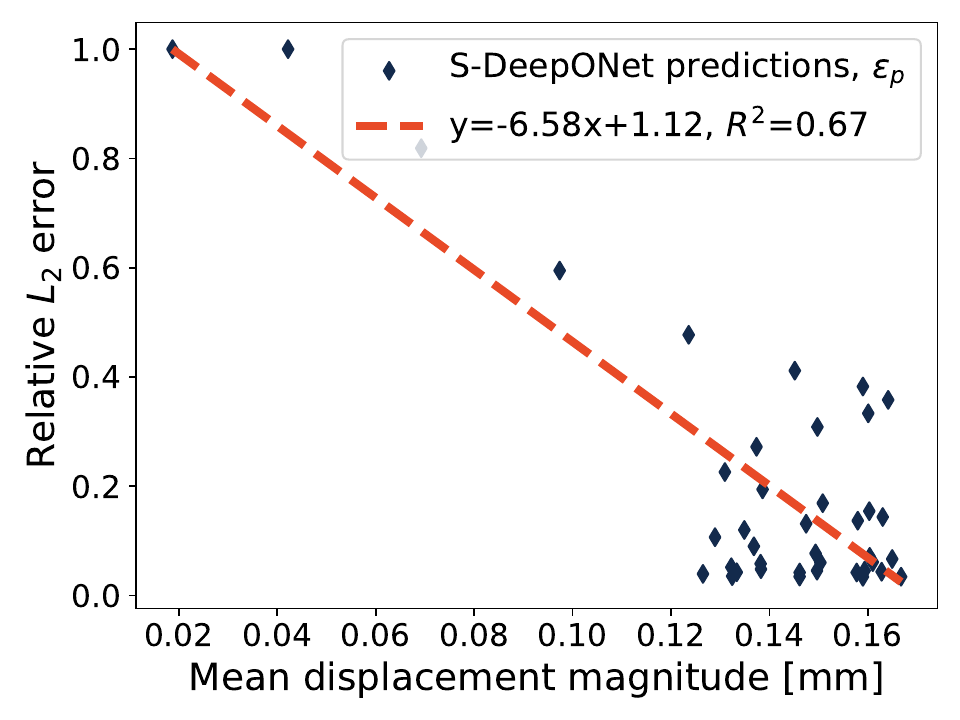}
         \label{lp2}
     }
    \caption{Scatter plots and trend lines of case-averaged errors for: \psubref{lp1} von Mises stress, \psubref{lp2} equivalent plastic strain.}
    \label{err_vs_u}
\end{figure}

We compare the histograms of the time-averaged error for von Mises stress and equivalent plastic strain for the vector and scalar S-DeepONets in \fref{model_comp_hist}. The scatter plots and trend lines for the prediction error versus the mean absolute magnitude of the applied displacement are shown in \fref{err_vs_u}.

From the repeatability results in \tref{cv1}, we see that the performance of the proposed vector S-DeepONet is consistent and repeatable, with minimal performance variation when trained with different data. With the von Mises stress, the model achieves a relative $L_2$ error of about 5\% and a $R^2$ value of 0.997, which is accurate considering that only 3200 data points were used in training. The relative $L_2$ error for equivalent plastic strain is as high as 21\%, but this is expected as this metric is ill-defined before the specimen yields (all elastic, no plastic strain). When inspecting the mean absolute error in plastic strain, we see that the error is only about $8.4\times10^{-3}$\%, and the $R^2$ value is 0.999, again indicating accurate model predictions. It is also worth highlighting that the proposed architecture solved two problems with completely different physics with only minor changes in network structures to account for the different number of time steps and output vector components, thereby demonstrating the generalizability and versatility of the proposed framework. The contour plots at different percentiles and time steps as shown in \fref{pct_plot1} provide a more direct view of the prediction performance. In the best case (first row of \fref{pct_plot1}), the S-DeepONet is able to accurately capture the hot spots in stress and plastic strains at different time steps, showing close agreement with the corresponding FE simulations. At the 80$^{th}$ percentile (second row of \fref{pct_plot1}), we see that the S-DeepONet was unable to predict the stress contour initially when the stress is small and response is fully elastic, but is still able to provide accurate predictions of both quantities at later time steps once the magnitudes of the stress and plastic strain increase. A similar trend is observed for the worst case (last row of \fref{pct_plot1}), where the S-DeepONet is unable to capture the stress distribution due to its small magnitude. Note that the specimen remains completely elastic throughout this loading history as evidenced from the reference FE simulation. The relation of prediction error and solution magnitude will be further explored later in the discussion.

The key improvement of the current architecture is the ability to predict multiple vector components at time steps with one model. For the case of two output components, we see from \tref{comp1} that the vector model only has 0.4\% more trainable parameters as compared to the scalar S-DeepONet model. Training the vector model took 15547s, which is 20.8\% faster than the combined training time for the two scalar models, while the inference speed is about 80\% faster. On average, each FE simulation requires 48s of CPU time to solve, making the S-DeepONet inference about 11900 times faster than running a FE simulation. Moreover, as revealed by the performance metrics presented in \tref{comp2}, we see that the performance of the vector S-DeepONet is highly similar to (albeit slightly worse) than the individual scalar networks. From \fref{model_comp}, we see that the predicted contour plots of the von Mises stress and equivalent plastic strain are highly similar at different time steps as well. Except at the beginning of the loading history (first column of \fref{model_comp}) where the specimen is fully elastic, the vector and scalar models predicted different contours for equivalent plastic strain (when they should be identically 0). \fref{model_comp_hist} compares the histograms of the time-averaged errors from the three different models. We see that the distribution of the prediction accuracy is similar between the vector and scalar S-DeepONet models. However, considering the significant time and model size savings, training the proposed vector S-DeepONet model instead of training two scalar models individually is the superior option from an efficiency perspective. 

Lastly, for a multi-step prediction, it is worth studying how the prediction error changes as a function of the mean magnitude of the applied displacement (which changes with time). From \fref{err_vs_u}, we see two slightly different behavior. For the von Mises stress (\fref{lp1}), the negative slope of the fitted trend line indicates that the error is, in general, decreasing with increasing load magnitude. However, the small $R^2$ value of the trend line indicates that the correlation is minimal. For equivalent plastic strain (\fref{lp2}), a similar decreasing trend is observed with increasing load magnitude, but this time with a much higher $R^2$ value. As a comparison, recall from \sref{sec:cfd} that all three components ($P,u$, and $v$) show a relative strong correlation with trend line $R^2 \ge 0.7$. This difference is reasonable since plasticity is path-dependent; the current stress magnitude depends not only on the current displacement magnitude but also on the past loading history, hence the lower correlation with the current displacement magnitude. On the other hand, the steady-state flow field can be uniquely defined by the lid velocity. Hence, the time-dependent flow field magnitude shows a stronger correlation with the current lid velocity. The limitation that the model cannot accurately predict the contour when the magnitude of the field variables is small is likely due to the use of the MSE loss function, where those data points only account for a small portion to the total loss and therefore not effectively minimized during training.

\subsection{Application of the trained S-DeepONet model}
\label{sec:trained}
\begin{figure}[h!] 
    \centering
     \subfloat[]{
         \includegraphics[trim={0.5cm 0.4cm 0.4cm .4cm},clip,width=0.31\textwidth]{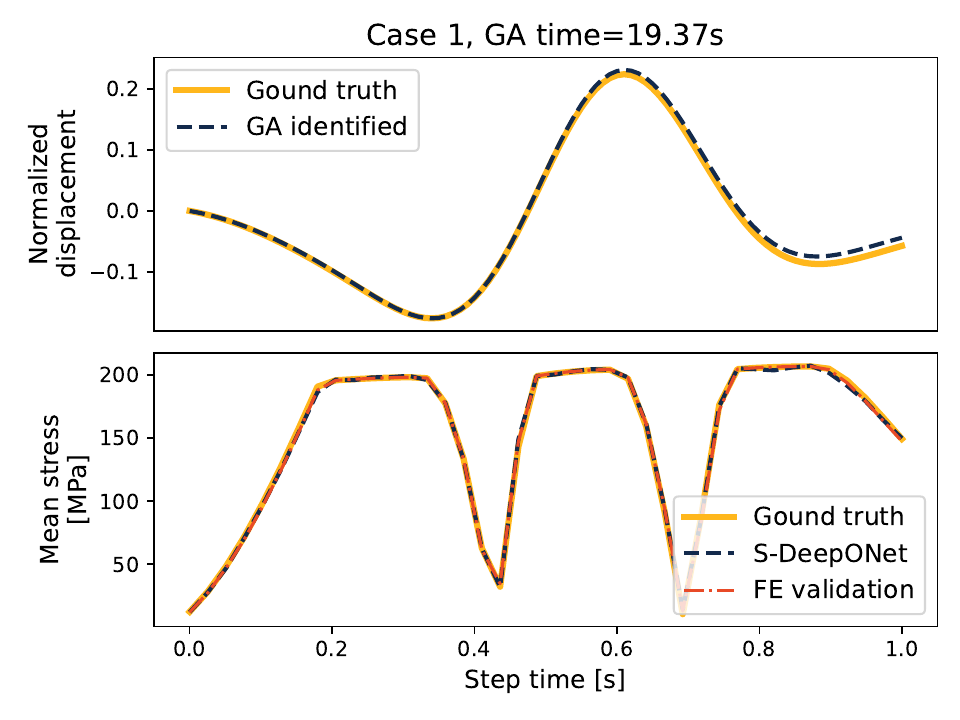}
         \label{iv1}
     }
     \subfloat[]{
         \includegraphics[trim={0.5cm 0.4cm 0.4cm .4cm},clip,width=0.31\textwidth]{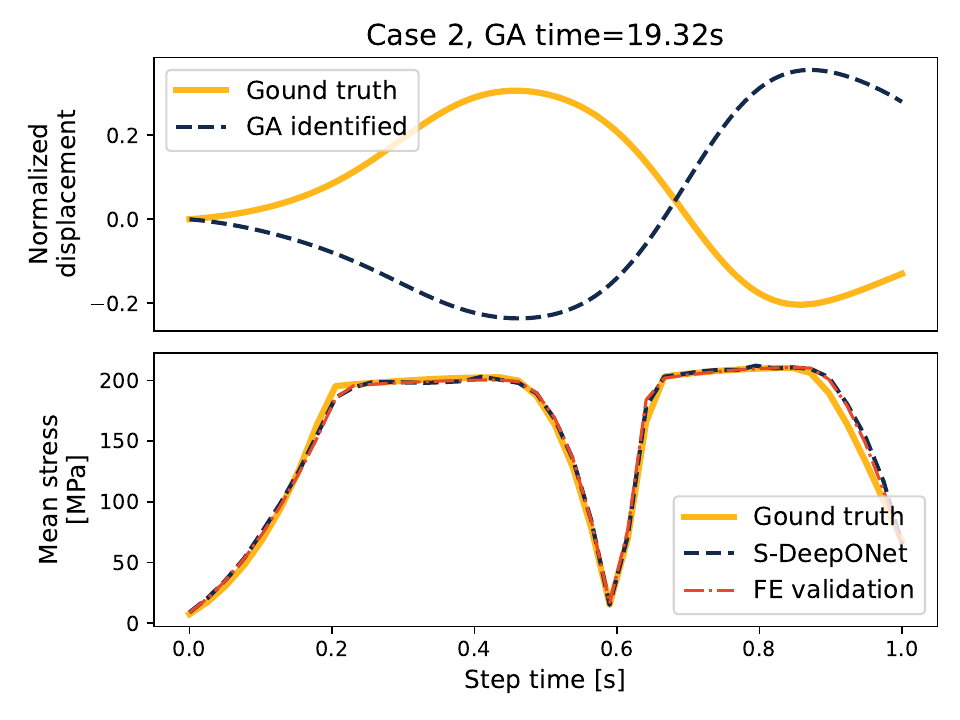}
         \label{iv2}
     }
     \subfloat[]{
         \includegraphics[trim={0.5cm 0.4cm 0.4cm .4cm},clip,width=0.31\textwidth]{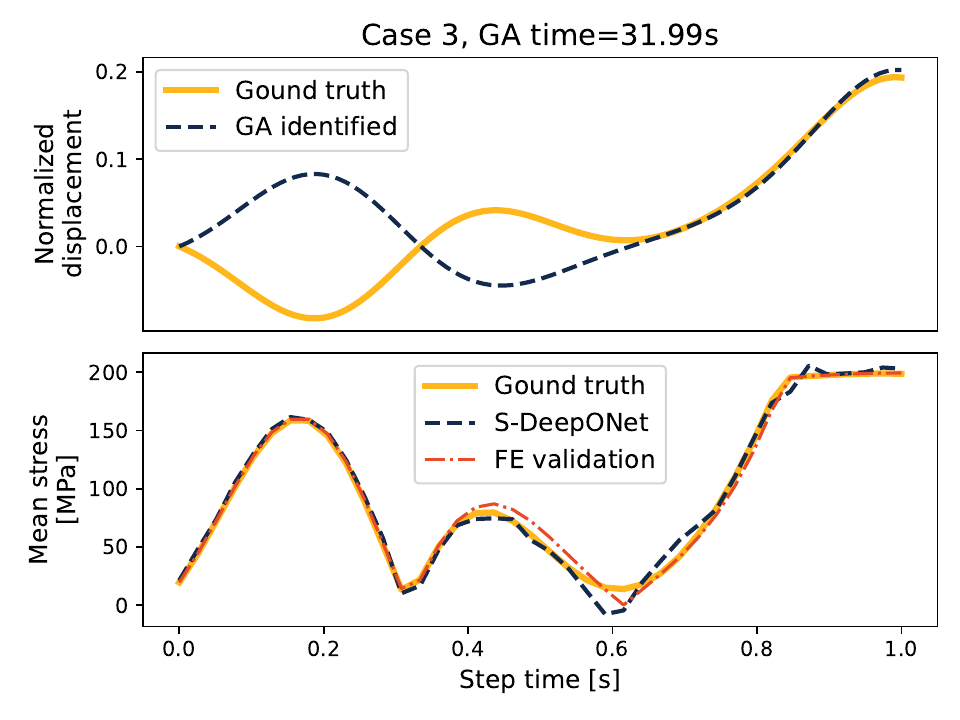}
         \label{iv3}
     }
    \caption{Comparison of the load curves and stress histories identified by S-DeepONet and GA.}
    \label{inv_comp}
\end{figure}

Three additional load curves were randomly generated following the same procedure outlined in \sref{sec:plastic_def}. Those curves were not seen by the trained S-DeepONet model during its training and testing. FE simulations were performed on those load curves to obtain the reference stress history curves, which were inputs to the inverse identification. To validate the results, FE simulations were performed using the load paths identified by S-DeepONet. The comparison of the load curves and stress histories are shown in \fref{inv_comp}.

With the help of efficient forward inference of the S-DeepONet model, the three GA cases were completed in an average of 24s, which is fast considering that a single forward FE simulation of the dog bone specimen takes around 48s. From the results, we see that in all three cases, the S-DeepONet predicted stress histories match closely with the given ground truths. Moreover, a comparison of the S-DeepONet predictions and the corresponding FE validation results (generated from the same load curve predicted by GA) shows that the S-DeepONet predictions remain highly accurate for load curves outside of the original training and testing data points, again showing the high accuracy of the trained model. Despite the high similarity between the identified and given stress histories, the load curves identified by GA and S-DeepONet do not always match the known ground truths. We see two characteristic cases from \fref{inv_comp}: (1) matching load curves (i.e., \fref{iv1}), (2) load curves with equal but opposite displacements (i.e., \fref{iv2}). It is also interesting to see a combination of both characteristic behaviors in \fref{iv3} where the second half of the predicted load curve matches closely with ground truth while the first half shows opposite displacements. This is reasonable from a mechanics perspective, since the von Mises stress is a positive quantity regardless of tensile or compressive loading, and the material considered in this work does not exhibit tension-compression asymmetry.

\section{Conclusions, limitations, and future work}
\label{sec:conc}
The sequential DeepONet (S-DeepONet) model previously proposed by the authors \cite{HE2024107258} is a variant of the DeepONet architecture \cite{lu2021learning} that uses a gated recurrent unit (GRU) in the branch network to capture the temporal information in the time-dependent input functions. However, the original S-DeepONet architecture only predicts the last time frame of a time-dependent evolution, and the predicted field is a scalar. Realizing this limitation, we introduced an improved version of the S-DeepONet capable of simultaneously predicting multiple time steps for solution fields with multiple vector components. This feature is made possible by the tensor-product structure of the combination operation that combines the encoded temporal information from the branch network and the encoded spatial information (for all vector components) from the trunk network. We further elucidate that the tensor-product combination process can be viewed as the simultaneous identification of a set of basis shapes and weights, with which the final contour can be expressed as a weighted linear combination. This architecture expands on the idea of exploiting the powerful temporal encoding capability of the GRU and the spatial encoding capability of the DeepONet architecture. To the best of the authors' knowledge, this is also the first time in literature that DeepONet is extended to predict a transient field with multiple vector components at different time steps. To demonstrate the application of the improved S-DeepONet, we showed an example of lid-driven cavity flow and an example of plastic deformation, both subjected to time-dependent input loads and having multiple output components at many time steps. In both cases, the improved S-DeepONet was able to provide accurate predictions for all output components using only 3200 data points in training. For all components, the DeepONet predictions achieved a $R^2$ value of over 0.99 and a relative $L_2$ error of less than 10\% (except for the equivalent plastic strain in \sref{sec:plastic_res}). We highlight the fact that minimum architecture change is needed (except for change the output components and time steps) for solving those two problems with totally different underlying physics, which shows that the proposed S-DeepONet is highly versatile. Once trained, the S-DeepONet can infer accurate full-field results at different time steps at least three orders of magnitude faster than direct numerical simulations. The trained model can also be used in conjunction with gradient-free optimizers such as the genetic algorithm to perform accurate inverse parameter identification. The example using the plasticity data shown that the inverse identification is highly efficient (finished in about half the time of one FE simulation) and accurate. Using the plasticity example, we also demonstrated the effectiveness of predicting all output components at once using a vector network instead of training a separate scalar network for each of the output component. With only 0.4\% more parameters, the vector S-DeepONet trained 20.8\% faster than the two scalar networks combined, while maintaining a very similar level of prediction accuracy. Therefore, it is recommended to use the vector S-DeepONet proposed in the current work, if multiple output components are of interest at multiple time steps. 

Through the analysis of the mean error magnitude as a function of the mean input load magnitude, it was revealed that the current S-DeepONet suffers from the limitation that it is unable to accurately predict the field contours when the field magnitude is small, even with a time-dependent data scaling scheme. This is likely due to the use of the MSE loss function, which only results in a small loss contribution for those data points with small magnitudes.

With the capability to accurately and efficiently predict solution history at multiple time steps for multiple different components, the improved S-DeepONet architecture proposed in this work provides a versatile tool to the engineering community to build surrogate models for complex nonlinear numerical simulations. The versatility of this architecture is demonstrated in this work through its application in both solid and fluid mechanics problems, and can be further employed in different engineering applications. With the weights and bias of the trained model, almost instant full-field forward predictions can be evaluated even on low-end platforms like laptops, which provides a novel tool for rapid preliminary designs, sensitivity analysis, uncertainty quantification, online controls, and as a black-box surrogate model for optimization, as we have demonstrated. In future work, we will investigate the use of transformer models \cite{jaderberg2015spatial} and attention mechanisms \cite{vaswani2017attention} in the DeepONet architecture to achieve higher prediction accuracy with lower computational costs.

\section*{Replication of results}
The data and source code that support the findings of this study can be found at \url{https://github.com/Jasiuk-Research-Group}. \textcolor{red}{Note to editor and reviewers: the link above will be made public upon the publication of this manuscript. During the review period, the data and source code can be made available upon request to the corresponding author.}

\section*{Conflict of interest}
The authors declare that they have no conflict of interest.

\section*{Acknowledgements}
The authors would like to thank the National Center for Supercomputing Applications (NCSA) at the University of Illinois, and particularly its Research Consulting Directorate, the Industry Program, and the Center for Artificial Intelligence Innovation (CAII) for their support and hardware resources. This research is a part of the Delta research computing project, which is supported by the National Science Foundation (award OCI 2005572) and the State of Illinois, as well as the Illinois Computes program supported by the University of Illinois Urbana-Champaign and the University of Illinois System. Finally, the authors would like to thank Professors George Karniadakis, Lu Lu, and the Crunch team at Brown, whose original work with DeepONets inspired this research.

\section*{CRediT author contributions}
\textbf{Junyan He}: Methodology, Formal analysis, Investigation, Writing - Original Draft.
\textbf{Shashank Kushwaha}: Methodology, Formal analysis, Investigation, Writing - Original Draft.
\textbf{Jaewan Park}: Methodology, Investigation, Writing - Original Draft.
\textbf{Seid Koric}: Conceptualization, Methodology, Supervision, Resources, Writing - Original Draft, Funding Acquisition.
\textbf{Diab Abueidda}: Supervision, Writing - Review \& Editing.
 \textbf{Iwona Jasiuk}: Supervision, Writing - Review \& Editing.

\bibliographystyle{unsrtnat}
\setlength{\bibsep}{0.0pt}
{\scriptsize \bibliography{References.bib} }
\end{document}